\documentclass[aps,prd,twocolumn,superscriptaddress,nofootinbib,eqsecnum,showpacs,letterpaper,11pt]{revtex4}

\usepackage{verbatim}
\usepackage[usenames,dvipsnames]{color}
\usepackage[pdftex]{graphicx}
\usepackage[english]{babel}
\usepackage{amsmath,amssymb}
\usepackage[colorlinks=true,citecolor=blue,linkcolor=magenta]{hyperref}
\usepackage{soul,color}

\begin{document}
\newcommand{\psiF}{\psi_{\omega}}
\newcommand{\psiFl}[1]{\psi_{\omega,#1}}
\newcommand{\sttheta}{\mbox{\st{$\theta$}}}
\newcommand{\E}{\mathrm{E}}
\newcommand{\Var}{\mathrm{Var}}
\newcommand{\bra}[1]{\langle #1|}
\newcommand{\ket}[1]{|#1\rangle}
\newcommand{\braket}[2]{\langle #1|#2 \rangle}
\newcommand{\mean}[2]{\langle #1 #2 \rangle}
\newcommand{\be}{\begin{equation}}
\newcommand{\ee}{\end{equation}}
\newcommand{\ba}{\begin{eqnarray}}
\newcommand{\ea}{\end{eqnarray}}
\newcommand{\SD}[1]{{\color{magenta}#1}}
\newcommand{\rem}[1]{{\sout{#1}}}
\newcommand{\alert}[1]{\textbf{\color{red} \uwave{#1}}}
\newcommand{\Y}[1]{\textcolor{yellow}{#1}}
\newcommand{\R}[1]{\textcolor{red}{{\it[#1]}}}
\newcommand{\B}[1]{\textcolor{blue}{#1}}
\newcommand{\C}[1]{\textcolor{cyan}{#1}}
\newcommand{\db}{\color{darkblue}}
\newcommand{\huan}[1]{\textcolor{red}{#1}}
\newcommand{\Marc}[1]{\textcolor{green}{#1}}
\newcommand{\intinfty}{\int_{-\infty}^{\infty}\!}
\newcommand{\Tr}{\mathop{\rm Tr}\nolimits}
\newcommand{\const}{\mathop{\rm const}\nolimits}
\newcommand{\zcoord}{Z} 
\newcommand{\phicoord}{\varphi} 
\newcommand{\obscoord}{\zeta} 
\newcommand{\src}{\mu} 
\newcommand{\mpart}{m} 
\newcommand{\indobs}{f} 
\newcommand{\indem}{i} 
\newcommand{\robs}{r_\indobs} 
\makeatletter
\newcommand{\rmnum}[1]{\romannumeral #1}
\newcommand{\Rmnum}[1]{\expandafter\@slowromancap\romannumeral #1@}
\makeatother
\newcommand{\fan}[1]{\textcolor{cyan}{#1}}

\newcommand{\CommMC}[1]{\textcolor{blue}{#1}}

\title{Wavefront twisting by rotating black holes: orbital angular momentum generation and phase coherent detection}
\author{Huan Yang}
\email{hyang@perimeterinstitute.ca}
\affiliation{Perimeter Institue for Theoretical Physics, Waterloo, Ontario N2L2Y5, Canada}
\affiliation{Institute for Quantum Computing, University of Waterloo, Waterloo, Ontario N2L3G1, Canada}
\author{Marc Casals}
\email{mcasals@cbpf.br}
\affiliation{Institute of Cosmology, Relativity and Astrophysics, Centro Brasileiro de Pesquisas F\'isicas, Rio de Janeiro, CEP 22290-180, Brazil.
}

\begin{abstract}
In this paper we study wave propagation and scattering near a black hole. In particular, we assume a coherent 
emission source near the black hole and investigate the wavefront distortion as seen by a distant observer. 
By ignoring the spin nature of the electromagnetic radiation we model it by a complex scalar field. Then, the propagating wave near the observer can be decomposed using the Laguerre-Gaussian mode basis and its wavefront distortion can be characterized by the decomposition coefficient. We find that this decomposition spectrum is symmetric
with respect to the azimuthal quantum number in the case that the wave source is located 
near a non-rotating (Schwarzschild) black hole, whereas the spectrum is generically asymmetric if the host black hole is rotating (Kerr).
The spectral asymmetry, or the net orbital angular momentum carried by the wave, is intimately related to the black hole spin and mass, the wave frequency and  the locations of the source  and the observer. We present semi-analytical expressions and numerical results for these parameter-dependences. 
If the emitted radiation is temporally coherent, our results show that the secondary images
(arising from the orbiting of the wavefront around the black hole) of the source can be almost as bright as its primary image.
Separately, in the case of temporally-incoherent radiation, we show that the non-fundamental spectrum components in the primary image
could be resolved by spatially-separated telescopes, although that would be degenerate
with the telescope direction.
Finally, our results suggest that the black-hole-induced spectral asymmetry is generally too weak to be observed in radio astronomy, 
even if the observer were located near an optical caustic.
\end{abstract}

\maketitle
%-----------------------------------------------------------------------------------------------------------------------------------------------------------------------------------------------------------

\section{Introduction}
Photon orbital angular momentum (POAM),
as compared to photon spin angular momentum, was less known in optics up until about two decades ago, mostly due to the technical difficulties in generating light with definite POAM states and in finding appropriate applications for such light. In 1990, Tamm and Weiss \cite{Tamm} first managed to produce Laguerre-Gaussian (LG) laser beams in the laboratory, which have helical phase front and quantized POAM\footnote{LG modes are spatial eigenmodes of a wave which is freely propagating under the paraxial approximation and which has integer orbital angular momentum. See Sec.~\ref{sec2} for further details.}. Their studies paved the way for later proposals on applications of LG modes, including applications on quantum information processing and quantum cryptography \cite{Mair, Leach, Terriza, Gibson}, or even on future generations of gravitational wave detectors \cite{Vinet}.

In addition, Harwitt \cite{Harwitt} proposed several astrophysical sources or mechanisms that possibly introduce nonzero POAM to light.
These sources and mechanisms  include maser beams that pass through inhomogeneous interstellar medium, luminous pulsars or quasars, and waves passing through the vicinity of a rotating black hole. Recently, Tamburini {\it et al.} \cite{Tamburini} performed a numerical simulation of radio emissions from an accretion disk surrounding a rotating black hole, assuming that different radiative sources in the disk are spatially coherent. In the simulation, they observed nontrivial POAM generation and asymmetric spectra in terms of
the  LG-mode basis, depending on the spin of the host black hole and the observer's  location in the sky. It remains physically important to understand the physical mechanism for the  generation of light with POAM near black holes, and obtain estimates for  the POAM magnitude, which apparently encodes information about the host black hole.  
  
 In this study we analyze the scalar
   wave emission from a coherent point source near a Kerr or a Schwarzschild black hole.
 We note that although the case of the electromagnetic wave which is considered in the above studies is a spin-$1$ field,
 in this paper we consider instead a complex scalar field, which has zero spin.
   We use the scalar field as a model for the electromagnetic field when its spin character is neglected.
 We employ this scalar model since it is a technically simpler case to study than the electromagnetic case and yet it
  is sufficient in order to understand the generation of POAM spectra.
  By assuming the wavelength of the  radiation (not greater than $mm$ scale) to be much smaller than the size of the black hole (not less than $km$ scale), we calculate the wave received by a distant observer on the celestial sphere and the corresponding POAM spectrum.
We investigate both the cases that the emission is temporally coherent and incoherent, as the two cases produce different POAM spectra. For completeness, we also study the scenario that the observer is located near an optical caustic of the background space-time, in which case the POAM asymmetry could be amplified, in addition to the wave itself. 

Part of the wave emitted from the vicinity of the black hole immediately propagates outwards and reaches the  far-away observer, thus yielding the so-called primary image.
However, other parts of the wave will typically orbit around the black hole a number of times before leaving the vicinity of the black hole and
propagating outwards to reach the  far-away observer; these
wave signals will correspond to secondary  images.
 We obtain the propagation of the wave via the calculation of an approximation to the retarded Green function of the wave equation in Kerr space-time.
 For the primary image we approximate the Green function by using the so-called  Hadamard form  (see, e.g.,\cite{Poisson}) and a calculation of the so-called van Vleck determinant,
 a biscalar which measures the degree of focusing of neighboring null geodesics.
 As for the later images -- for which the Hadamard form is not valid -- we instead approximate the Green function by a calculation of the quasi-normal modes in Kerr space-time
 in the high-oscillation-frequency limit (see, e.g.,\cite{Yang2013}).
 
     We shall show that, although the emission from a single non-rotating star in flat space-time generally contains only the fundamental LG mode
     (which contains zero angular momentum) at far distances, the presence of a rotating black hole near the star will generate a non-trivial POAM spectrum.
    Such spectrum is symmetric with respect to the LG basis (i.e., with respect to the azimuthal quantum number) 
    for non-rotating Schwarzschild black holes and generically asymmetric for rotating Kerr black holes.
That is, the asymmetric part of the spectrum contains the spin information of rotating black holes.
As we shall show in Sec.~\ref{sec:specdegen}, the symmetric part of a POAM spectrum may be affected by the direction of the observation plane of a telescope array.
This further emphasizes the importance of measuring the asymmetric part of the spectrum.

An important difference between the case we study in this paper and that in \cite{Tamburini} is that here, as opposed to \cite{Tamburini},
we have 
a pointlike
emission source. Therefore, the interference between waves emitted from sources at different spatial locations that occurs in \cite{Tamburini} is absent here.
In our case, the main effect comes only from gravitationally twisting/merging the light bundles from a single emission source and we therefore expect
 the spectral asymmetry  to be much smaller than in  \cite{Tamburini} (see Secs.~\ref{sec:wave} and \ref{sec4} for details).

This paper is organized as follows. In Sec.~\ref{sec2} we review the decomposition of a paraxial wave with respect to the LG basis, the definition of a POAM spectrum and the related quantity for detection.
In Sec.~\ref{sec:GF} we describe the methods used for the calculation of the Green function.
 In Sec.~\ref{sec:wave} we analyze the wave emitted by a source near a black hole, using the Green function approach, and present our POAM results. In Sec.~\ref{sec4}
 we investigate the setting where the observer is located near an optical caustic and we conclude in Sec.~\ref{sec5}.
Throughout this paper, we use geometric units $G=c=1$,
the black hole mass $M$ is also set to $1$, unless otherwise specified, and
the 
metric signature is taken to be $(- + + +)$.

  %-----------------------------------------------------------------------------------------------------------------------------------------------------------------------------------------------------------

   \section{Overview of POAM and spectral decomposition} \label{sec2}
  
  In this section, we present a cursory review of the LG modes of light and the corresponding POAM-spectrum decomposition, which is useful for later sections. Interested readers may find a detailed discussion in \cite{Elias} on topics such as POAM observables in astronomy, the propagation or map of POAM from the celestial sphere onto detectors, the detection of POAM using existing astronomical instruments, etc.
  
  \subsection{Mode decomposition} \label{sec:mode decomp}

Let us consider a light beam propagating on a general space-time.
    We now describe two approximations. We will apply the first  to the propagation of the beam all the way  
     from the source to the observer (i.e., throughout the curved space-time), while will apply  the second one only near the observer (i.e., where the space-time
     is asymptotically flat).
     
 The first approximation is the `scalar field approximation', under which the vector character of the electromagnetic field is neglected.
  Under this approximation, the electromagnetic field is just described by a distribution of the field amplitude and phase, which is given by a
  complex scalar wave function  $\Psi(x)$ that satisfies the Klein-Gordon equation:
 \begin{align} \label{eq:KG}
\Box \Psi(x)
&=\frac{1}{\sqrt{-g}}\partial_{\mu}\left(\sqrt{-g}g^{\mu\nu}\partial_{\nu}\Psi(x)\right)
\notag \\ &
=-4\pi \src(x),
\end{align}
where  
$x$ is a space-time point,
$g_{\mu\nu}$ is the background space-time metric, $g(x)\equiv{\rm det}(g_{\mu\nu}(x))$ and
$\src(x)$ is the source of the field.
Physically, this approximation is valid, for example, when one of the components of the traverse field dominates over the other one. 
In flat spacetime (such as near the observation plane in an astrophysical context), where  it is a common approximation (e.g.,
\cite{Elias,Thorne}), one may alternatively view
the scalar field $\Psi$ as representing one of the Cartesian components of the electric or magnetic field.
In curved space-time, we expect this approximation to be  valid in the he high-frequency limit that we adopt in this paper, particularly in the situation of
Sec.\ref{sec3b} where only the part of the wave on, or closely near, one single image (the primary one) is considered.
In the situation of Sec.\ref{sec3a}, where various images -- which  folloow different paths around the black hole -- are considered this approximation may be less 
valid, because the polarization of the electromagnetic field could rotate differently along different path. 
In that case, more careful analysis has to be done to take into account the rotation of field polarizations.

Let us now carry out a Fourier-mode decomposition of the complex scalar field $\Psi(x)$
   and  denote its 
Fourier modes by  $\psiF=\psiF(\bf x)$, where $\bf x$ denotes the spatial coordinates of $x$ and $\omega$ the frequency.
Since in this paper we will be considering a source radiating with a single frequency, the spatial part of the 
total field $\Psi(x)$ will be given by a single Fourier mode (see Eq.(\ref{eqpsiGreen}) below; the only exception will be in Eq.(\ref{eqmulfreq})).

In order to introduce the LG mode decomposition now and for the rest of this subsection we assume flat space-time.
Even though in this paper we deal with the presence of a black hole, the emmitted wave will propagate from the strong-field near the black hole all the way
to Earth, where the space-time is asymptotically flat. Here, a connection can be made between the emitted wave as detected on Earth and the LG basis
defined in flat space-time. 
In flat space-time, the Fourier modes  satisfy the Helmholtz equation:
 \begin{equation} \label{eq:Helmholtz}
  (\nabla^2+\omega^2)\psiF=0\,,
  \end{equation}
  where $\omega=||\vec{k}||$, $\vec{k}$
  is the wave vector
   and $\nabla^2$ is the Laplacian operator in flat space-time.

The LG basis requires a second approximation:  the paraxial approximation, under which the wave propagates approximately along an axis, 
   say the $\zcoord$-axis
  (i.e., $k_X$,$k_Y\ll k_{\zcoord}\approx  ||\vec{k}||$ using Cartesian coordinates). This is generally a good approximation when the light
(even though strictly speaking we are now dealing with a scalar field, we use it to model light -- within the scalar field approximation --
  and so at times we might still  refer to the wave as `light') is far away from its emission source.
  This approximation is well justified in the astrophysical setting when the wave has reached a distant observer after travelling 
  all the way   from  near the black hole and  so  the celestial sphere is locally a plane.
 Under the paraxial approximation the spatial wave function $\psiF$ can be expanded as a family of LG modes $u_{pl}$,
  which have the following form (e.g.,~\cite{Siegman}):
  \begin{widetext}
  \begin{equation}\label{eqlgwf}
u_{pl}(\zcoord,\rho,\phicoord)=\frac{C}{w(\zcoord)}\left (\frac{\rho \sqrt{2}}{w(\zcoord)}\right)^{|l|}e^{-\rho^2/w(\zcoord)^2}L^{|l|}_p\left(\frac{2\rho^2}{w(\zcoord)^2}\right)e^{i \omega \rho^2/(2R(\zcoord))}e^{i l\phicoord}e^{i(2p+|l|+1)\xi(\zcoord)+i \omega \zcoord},
\end{equation}
\end{widetext}
where 
   we use a cylindrical coordinate system $(\zcoord,\rho,\phicoord)$ for the spatial point $\bf x$,
$C$ is a normalization constant, $L^{|l|}_p$ is a Laguerre polynomial, $l$ is the azimuthal quantum number and $p$ is the radial quantum number.
We have  defined the functions
\begin{align}
w(\zcoord)&\equiv w_0\sqrt{1+\zcoord^2/z^2_R},\, R(\zcoord)\equiv \zcoord\left ( 1+\frac{\zcoord^2_R}{\zcoord^2}\right ),\nonumber \\ 
\xi(\zcoord)&\equiv\arctan(\zcoord/\zcoord_R),
\end{align}
where $w_0$ is the size of the beam waist, $\zcoord_R\equiv \pi w^2_0/\lambda$ is the Rayleigh length and $\lambda=2\pi/k$  is the wavelength of the light.

Because of the $e^{i l \phicoord}$ factor, the phase fronts of LG modes (except for the fundamental mode $l=0$) are helical, and the mode with indices $\{l, p\}$ carries a definite orbital angular momentum $l \hbar$ per photon. This can be intuitively understood by noticing that the wave vectors are spiraling around the propagation ($\zcoord$) axis ($k_x$ and $k_y$ are small
but non-zero). It can also be shown that LG modes form a complete basis for all the paraxial waves with the same frequency.
Therefore, at a given value of $\zcoord $, any spatial wave function $\psiF$ can be decomposed in the basis of LG modes as
\begin{equation}\label{eq:LG basis}
\psiF (\zcoord, \rho, \phicoord)= \sum_{p = 0}^{\infty}\sum_{l= -\infty}^{\infty} c_{pl}(\zcoord) \,u_{pl}(\zcoord, \rho, \phicoord) \,,
\end{equation}
where the coefficients are given by
\begin{equation}
c_{pl}(\zcoord) = \langle u_{pl} | \psiF \rangle \,,
\end{equation}
 where the product of any two spatial functions $u$ and  $\psiF$ is defined on the constant-$\zcoord$ plane as:
 \begin{equation}
 \langle u |\psiF \rangle \equiv \int^{2\pi}_0d\phicoord \int ^\infty_0 d\rho \,\rho\,  u^*(\zcoord, \rho, \phicoord)  \psiF (\zcoord, \rho, \phicoord)\,.
 \end{equation}

  For the purpose of understanding the generation of POAM spectra, 
  there is limited interest in distinguishing the radial profile of the wave function. As a result, we sum up all the radial components for each
  value of $l$, so that the decomposition becomes
 \begin{equation}
 \psiF(\zcoord,\rho,\phicoord) = \sum_{l=-\infty}^{\infty}  e^{i l \phicoord}\psiFl{l}(\zcoord,\rho )\,, 
  \end{equation}
  where $\psiFl{l}$ is given by
  \begin{equation}\label{eqandec}
 \psiFl{l}(\zcoord,\rho)=\frac{1}{2\pi} \int^{2\pi}_0 d\phicoord e^{-i l \phicoord} \psiF(\zcoord,\rho,\phicoord)\,.
  \end{equation}
  
    We further define the spectra weight $w_l$ of the POAM decomposition as
  \begin{align}\label{eqweight}
  w_l &=\frac{2\pi}{I} \int^\infty_0 d\rho \rho |\psiFl{l}|^2 \,, \\
  I & \equiv \int^\infty_0 d\rho\int^{2\pi}_0 \rho d\phicoord |\psiF|^2 \,.\label{eqnorm}
  \end{align}

  Based on their definition, it is easy to verify that the weights $w_l$ satisfy the normalization condition $\sum_l w_l =1$, and so the averaged POAM per photon is given by
  \begin{equation} \label{eq:POAM}
  \bar{L}_{\zcoord} = \hbar \sum^\infty_{l=-\infty} l\,w_l\,.
  \end{equation}
  
  The definition of $w_l$ can easily be generalized \cite{Elias} to incoherent light by decomposing the wave function as a summation of coherent pieces and dropping the mutual-interference terms.
Its physical properties can be further elaborated by studying the two following simple examples. 
   
 \subsection{Examples}  
 In the first example, we note that a plane wave, with its propagation direction orthogonal to the observation plane, satisfies $w_l =\delta_{l0}$, i.e, all modes have zero spectra weight except for the fundamental mode. This fact is important as any initially distorted wavefront always tends to flatten out during propagation in free space, because of the diffraction effect. Since real telescopes have a finite collection area, which also means that 
 the integration upper bound for $\rho$  in Eqs.~(\ref{eqweight}) and (\ref{eqnorm}) should be replaced by the instrument's size, the spectra weights for the modes with $|l|>0$  always decrease as we increase the distance between the detector and the  source.
  
\begin{figure}[t,b]\centering
\includegraphics[width=0.9\columnwidth]{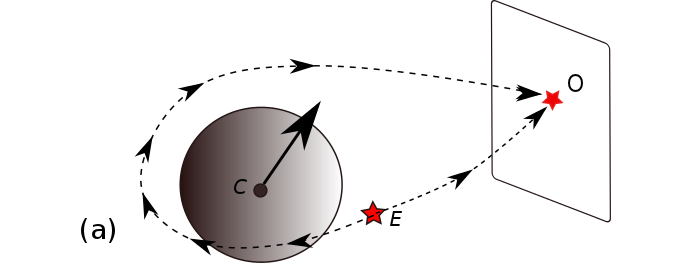}\\
\includegraphics[width=0.9\columnwidth]{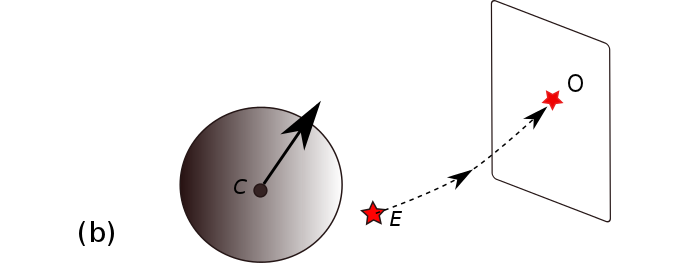}\\
\includegraphics[width=0.9\columnwidth]{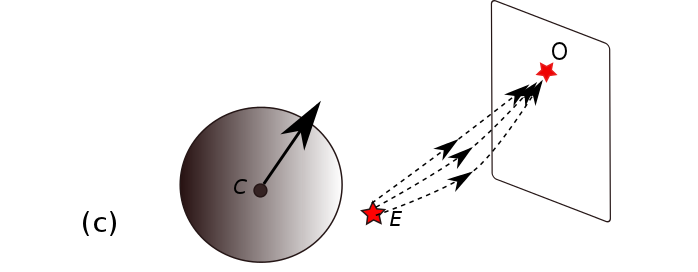}
\caption{(Color online.) 
An illustration of the three scenarios discussed in Sec.~\ref{sec3a}, Sec.~\ref{sec3b} and Sec.~\ref{sec4} respectively.  (a) The interference between two distinct rays that emit from E and meet each other at the observer O. (b) No interference between different rays. The light bundle of the primary ray is slightly twisted by the rotating black-hole spacetime. (c) Interference between adjacent light bundles. This only happens when the observer O is located near an optical caustic of the emission from E.} 
\label{fig:interf}
\end{figure}

 In the second example, if the wave function has a dependence $\psiF = e^{i \Phi(\zcoord,\rho,\phicoord)}$, where $\Phi(\zcoord,\rho,\phicoord)$ is a real-valued function, it can be shown that the $w_l$'s are generically nonzero and the spectrum must be symmetric, i.e., $w_l=w_{-l}$ (see Sec.~\ref{sec3b} for a proof in the case $l=\pm 1$; the generalization to higher $|l|$ is trivial). In fact, if we draw the normal vectors to the constant-phase plane of a LG mode with $l \neq 0$, it shows a spiraling pattern which geometrically resembles a twisted light bundle. This twisting of light can be generated, for example, by the frame-dragging caused by a rotating black hole (Sec.~\ref{sec3b}), by merging adjacent light bundles near an optical caustic (Sec.~\ref{sec4}), or even by the offset-interference between previously far-apart geometric rays (Sec.~\ref{sec3a}) --  see Fig.~\ref{fig:interf}. In all these scenarios, we observe variations in both the phase and the amplitude. This is also true for the LG modes as shown in Eq.~(\ref{eqlgwf}), where we find zero intensity at the origin ($\rho=0$) and nonzero amplitude elsewhere.

 When dealing with actual observables, it is more convenient to use $\sqrt{w_l}$ rather than $w_l$, because $\sqrt{w_l}$ is directly proportional to the wave amplitude at each $l$, which is an observable obtainable by Eq.~(\ref{eqandec}) or physically using an antenna array as illustrated in \cite{Thide}. As a consequence, although we shall refer to $|\sqrt{w_l}-\sqrt{w_{-l}}|$ as the spectral asymmetry throughout this paper, more strictly speaking it corresponds to the ``amplitude" asymmetry. This ``amplitude" asymmetry can be measured by comparing phases on different sites of the telescope array, and its detectability is closely related to the phase sensitivity of the telescope array. The ground-based radio telescopes suffer from  air-turbulence-induced phase errors, and by implementing novel techniques, such as measuring the {\it closure phase} \cite{Rogers}, the phase sensitivity can be improved to a scale of multiple wavelengths  \cite{Averycommu}.

 \subsection{Spectral degeneracy}\label{sec:specdegen}
 
A symmetric weight spectrum
  can be  generated by tilting the observation plane with respect to the optical axis, as this generally introduces a $e^{i\omega \robs (\alpha x_1+\beta x_2)}$
   factor into the wave, depending on the tilt angles $\alpha$ and $\beta$,
   where $\robs$ is the radius between the source (or approximately the black hole) and the observation plane, and
   $x_{1,2}$ are coordinates on the observation plane defined in Sec.\ref{sec3a}. Even for an incoming plane wave, this phase factor will generate a {\it symmetric} POAM spectrum, with $w_1=w_{-1} \propto \omega^2 b^2(\alpha^2+\beta^2)$, where  $b$ is the size of the detector.
   Unless the tilt angles are known within an accuracy of $M/\robs$, which is an unrealistic requirement, it generates a
   much larger symmetric POAM spectrum than the $l \neq 0$ components of the POAM spectrum induced by a rotating black-hole geometry.
Thus, the physics carried by a symmetric POAM is degenerate with the tilt angle of the observation plane, and is not likely to be decoded  via radio observations alone.
 
%-----------------------------------------------------------------------------------------------------------------------------------------------------------------------------------------------------------
 
 \section{Green function method}\label{sec:GF}
 
 In order to determine the wave emission by a point source, we shall 
 make use of the retarded Green function of the wave equation (\ref{eq:KG}) satisfied by the
 field propagating in the background space-time.
 The retarded Green function can also be used to calculate the self-force on a small compact object moving on the background of a massive black hole 
 (see, e.g.,~\cite{Poisson}) and the associated radiation back-reaction.
Techniques for the calculation of the Green function for field perturbations of different spin 
 in Schwarzschild, Kerr or other more exotic background space-times have been recently 
 developed~\cite{WardellOtt, Casals2009, Dolan2011, anil,  Casals2012a, Yang2013} and applied~\cite{Casals2009a,Casals2013,wardell2014self}
 with  success to the calculation of the self-force.
 In this paper we will employ some of these techniques for the calculation of the Green function to calculate the propagation of a scalar wave in Kerr space-time
 and we refer the reader to this literature for further details on the techniques.
 
 As explained in Sec.(\ref{sec:mode decomp}), we use 
 a massless, complex scalar field  $\Psi(x)$ to model the electromagnetic field when  its spin character is neglected.
This scalar field propagating on a curved space-time obeys the Klein-Gordon equation (\ref{eq:KG}).
 The retarded Green function satisfies this equation with an invariant Dirac-delta distribution in the source:
 \begin{align}
\label{scalargreen}
\Box G_{\rm ret}(x,x')
=-4\pi\frac{\delta_{4}(x-x')}{\sqrt{-g(x)}},\,
\end{align} 
subject to
 causal boundary conditions: 
 $G_{\rm ret}(x,x')$
 is zero  if $x$ does 
not lie in the causal future of $x'$. 
The propagation of the massless scalar wave
is determined by the retarded Green function via
\begin{equation} \label{eq:psi via GF}
\Psi(x)=\int d^4x''\sqrt{-g(x'')}\mu(x'')G_{\rm ret}(x,x'').
\end{equation}
From now on we will focus on the case that the background space-time  is Kerr space-time.
Let us assume that there is an emitter following a worldline given by $x'(\tau)$, where $\tau$ is the  proper time of the emitter, which is located near a rotating Kerr black hole and is radiating scalar waves with frequency $\omega_0$ with respect to its proper reference frame. 
Specifically, we take the source to be
\begin{equation}
\src(x)=\int_{-\infty}^{\infty}d\tau \frac{\delta_4(x-x'(\tau))}{\sqrt{-g(x)}}e^{-i\omega_0 \tau}.
\end{equation}
From Eq.(\ref{eq:psi via GF}), the wave that an observer
 located at the space-time point $x$ receives is 
 simply given by 
\begin{equation}\label{eqconv}
\Psi(x) =\int^\infty_{-\infty} d\tau e^{-i \omega_0 \tau} G_{\rm ret}\left(x, x'(\tau)\right)\,.
\end{equation}
Although the integration upper bound is here taken to be $\infty$, the integration actually contains only the causal contributions because the retarded Green function is zero if $x$ does not lie in the causal future of $x'$. For simplicity,  we assume hereafter that the emitter stays at a fixed spatial location $\bf x'$ at all times. Despite the fact that this is not a very physical scenario, it does not harm the main physical results and it simplifies the calculations.
We define the time-dilation factor $N$ and the rescaled frequency $\omega$ to be
\begin{equation}\label{eqNconven}
N \equiv \sqrt{-g_{tt}(\bf x')}\,,\quad \omega \equiv N \omega_0\,.
\end{equation}

 We also note that the time-translational symmetry of Kerr space-time implies
  $G_{\rm ret}(x, x')=G_{\rm ret}({\bf x},{\bf x'};t-t')$ 
  (with a slight abuse of notation).
 Using the fact that the proper time of the static emitter satisfies $d\tau =N dt'$ and plugging Eq.~(\ref{eqNconven}) into Eq.~(\ref{eqconv}), we obtain
 \begin{equation}\label{eqpsiGreen} 
 \Psi(x) =e^{-i \omega t} \psiF({\bf x}), \quad \psiF({\bf x})= N  G^\omega_{\rm ret}(\bf x, \bf x'),
 \end{equation}
 where $G^\omega_{\rm ret}$ are the Fourier modes of the retarded Green function.

 In the following subsections we will describe 
 two different techniques for calculating the Fourier modes
 $G^\omega_{\rm ret}$, 
 one technique being valid for the calculation of the primary image
 and the other for the calculation of the later, secondary images.
 Even though we develop the techniques for general Fourier frequencies, at some point in the analysis we will carry out approximations corresponding to the 
 high-frequency limit for the radiation (i.e., $\omega \gg 1/M$) which, as mentioned in the introduction, is the physical limit of interest in this paper.
 
%---------------------------------------------------------------------------------------------
 
 \subsection{Primary image: Hadamard form}\label{sec:primary image}

  In a `sufficiently' small neighborhood called a  normal neighborhood
\footnote{ 
A normal neighborhood of $x'$ is a region containing $x'$ such that every point $x$ in this region is connected to $x'$ by a unique geodesic in that region.}
of the point $x'$ of 
emission
of the primary pulse, the retarded Green function can be calculated via the  Hadamard form \cite{Poisson, Friedlander}
\footnote{In principle, it should also be possible to express the Green function in the normal neighborhood via the spectroscopic decomposition of Eq.~(\ref{eqgreenexpansion}). However, in practice the convergence of $G_{\rm QNM}$ and $G_{\rm BC}$ is usually poor at very early times \cite{Casals2012c}, and it is preferable to use the Hadamard form instead.},
\begin{align}\label{eqhada}
G_{\rm ret}(x, x') =[U(x,x')\delta(\sigma(x,x'))+ \nonumber \\ V(x,x')H(-\sigma(x,x'))]H_+(x,x').
\end{align}
Here, $U(x,x')$ and  $V(x, x')$ are both smooth biscalars, 
$H(x)$ is the Heaviside distribution
and $H_+(x, x')$ equals $1$ if $x$ lies to the future of $x'$ and equals $0$ otherwise. Synge's world function $\sigma(x, x')$ is given by
\begin{equation}\label{eqsigma}
\sigma(x,x')=
\frac{1}{2}(\bar\lambda_\indobs-\bar\lambda_\indem)\int^{\bar\lambda_\indobs}_{\bar\lambda_\indem} d\bar \lambda\, g_{\mu\nu}(\bar z) \,t^{\mu} t^{\nu}\,
\end{equation}
where the integration is performed along the unique geodesic 
$\bar z(\bar \lambda)$ 
connecting $x'$ and $x$, with
$\bar \lambda$ being an affine parameter and 
 $\bar z(\bar\lambda_\indem)=x'$ and $\bar z(\bar\lambda_\indobs)=x$.
Based on its definition, it is immediate to find that $\sigma$ is positive if the geodesic joining $x$ and $x'$ is spacelike, negative if it is timekike and zero if it is null.
Consequently, the term  `$U\delta(\sigma)H_+$' in Eq.(\ref{eqhada}) only has support along the `direct' null wavefront that propagates from $x'$ to $x$ and so we will refer
to it as the `direct part'.
We will refer to the points where this direct part of the null wavefront has support as the (future) light-cone.
The direct part of the Hadamard form, therefore, describes the primary image of a source at $\bf x'$ that an observer at $\bf x$ will detect.

The Fourier modes $G_{\rm dir}^{\omega}$ of the  direct part of the Hadamard form
correspond to the leading-order of the Fourier modes of the full Green function
in the geometrical optics limit
(or under the WKB approximation). Thus, according to \cite{anil} they are given by
  \begin{equation}\label{eqdirg}
 G_{\rm dir}^{\omega}({\bf x}, {\bf x'}) =
 A_0({\bf x}, {\bf x'})
 e^{i \,\omega T_0({\bf x}, {\bf x'})},
 \end{equation}
 where $\omega T_0(\bf x, \bf x')$ is the geometric phase for the `direct' null geodesic connecting $x$ and $x'$. The amplitude $A_0(\bf x, \bf x')$ is discussed in \cite{anil} in the context of mapping the areas of light bundles in a general curved space-time.
In the remaining of this subsection  we focus on the calculation of $A_0({\bf x}, {\bf x'})$ and $T_0({\bf x}, {\bf x'})$ in Kerr space-time.

By taking the Fourier transform of  the direct part  in Eq.(\ref{eqhada})  and comparing it with Eq.(\ref{eqdirg}), it is straightforward to find that
\begin{equation}\label{eq:direct part equality}
A_0({\bf x}, {\bf x'}) = \frac{U(x',x)}{\left |{\partial \sigma}/{\partial t} \right |_{t=T_0}}\,,
\end{equation}
and $\sigma=0$ when $t-t'=T_0({\bf x}, {\bf x'})$.
In order to evaluate $\partial \sigma/\partial t$, we use a family of timelike geodesics for a particle with rest mass $\mpart \to 0$ to asymptote to the null geodesic.
For timelike geodesics, we take the affine parameter $\bar\lambda$  to be  the proper time $\bar\tau$ along the timelike geodesic connecting $x'$ and $x$, with
 $\bar z(\bar\tau_\indem)=x'$ and $\bar z(\bar\tau_\indobs)=x$.
 Therefore, for timelike geodesics it is $\sigma =-(\bar\tau_\indobs-\bar\tau_\indem)^2/2$ and we have
\begin{align} \label{eq:dsigma/dt}
\left.\frac{\partial \sigma}{\partial t}\right|_{t=T_0} &=-\lim_{\mpart \to 0} (\bar\tau_\indobs-\bar\tau_\indem) \left .\frac{\partial  \bar\tau}{\partial t} \right |_{t=T_0, {\rm fixed }\, {\bf x'}}  \nonumber \\
&=- \lim_{\mpart \to 0} (\bar\tau_\indobs-\bar\tau_\indem) u_t \nonumber \\
& =  \lim_{\mpart \to 0}(\bar\tau_\indobs-\bar\tau_\indem)\frac{E}{\mpart} \nonumber \\
& = \bar \lambda_\indobs-\bar \lambda_\indem \,.
\end{align}
Here, $u_t=\partial_t \cdot u = -E/m$, where $u$ is the $4$-velocity, and $E$ is the energy of the massive particle. Geodesics for any massless particle can be obtained by integrating out the equations of motion, which are separable in Boyer-Lindquist coordinates $\{r,\theta,\phi,t\}$ in Kerr space-time and they are given by \cite{mtw}:
 \begin{align}
 \frac{d t}{d \bar\lambda} &= \frac{\left [ -a(aE\sin^2\theta-L_z)+(r^2+a^2)\Delta^{-1} P\right ]}{r^2+a^2\cos^2\theta} \,,\nonumber \\
 \frac{ d \phi}{d \bar\lambda} &= \frac{\left [ -(aE-L_z/\sin^2\theta)+a\Delta^{-1} P\right ]}{r^2+a^2\cos^2\theta} \,,\nonumber \\
  \frac{ d \theta}{d \bar\lambda} &= \frac{\left [ Q-\cos^2\theta(-a^2 E^2+L^2_z/\sin^2\theta)\right ]^{1/2}}{r^2+a^2\cos^2\theta} \,,\nonumber \\ 
  \frac{ d r}{d \bar\lambda} &= \frac{\left \{ P^2-\Delta[Q+(L_z-a E)^2]\right \}^{1/2}}{r^2+a^2\cos^2\theta} \,.
  \label{eq:NG}
\end{align}
Here, $E$ is the energy of the massless particle, $L_z$ is the angular momentum of the particle along the $z$-axis (the symmetric axis of the black hole; not to be confused
with the $Z$-axis, which is the axis of propagation of the wave used in Sec.\ref{sec2}; similarly, here $\phi$ is the azimuthal angle with respect to the symmetry axis of the black hole
whereas $\phicoord$ used in Sec.\ref{sec2} was the azimuthal angle with respect to the axis of propagation of the wave)
 and $Q$, which shows up in the expressions for $d\theta/d\bar\lambda$ and $dr/d\bar\lambda$, is  Carter's constant of  motion. The functions $\Delta$ and $P$ are given by
\begin{align}
\Delta \equiv r^2-2 M r+a^2,\quad P\equiv E(r^2+a^2)-L_z a\,,
\end{align}
where $M$ and $a$ are, respectively, the mass and the angular momentum of the Kerr black hole.
We integrate the set of coupled, first-order ordinary differential equations Eq.(\ref{eq:NG}) in order to obtain 
$(\bar\lambda_\indobs-\bar\lambda_\indem)$ (and therefore $\left.\partial \sigma/\partial t\right|_{t=T_0}$)
along the desired `direct' null geodesic $\bar z(\bar\lambda)$.
Further below (namely in order to calculate $\left .\partial T_0/\partial \obscoord\right |_{\bf x_0}$ in Eq.(\ref{eqpsi1coh})), we will also need the values of
$\partial T_0/\partial \theta_\indobs$ and $\partial T_0/\partial \phi_\indobs$, where $\theta_\indobs\equiv \theta(\bar\lambda_\indobs)$ and $\phi_\indobs\equiv \phi(\bar\lambda_\indobs)$.
In order to calculate these partial derivatives, we give small variations to $E$ and $L_z$ and find how $T_0$, $\theta_\indobs$ and $\phi_\indobs$ vary and we then  apply the chain rule. 
We note that these variations in $\theta_\indobs$ and $\phi_\indobs$ allow us to find (an approximation to) the Jacobian 
$\frac{\partial\{\theta_\indobs,\phi_\indobs\}}{\partial\{E,L_z\}}$, which allows us to calculate the values of $E$ and $L_z$ required in order for the final spatial point $\bf x$ to remain
the same as we vary the black hole angular momentum $a$.

As for the bitensor $U(x, x')$, in four-dimensional space-times it is related to the van Vleck determinant $\Delta_{\bar z}(x,x')$  as $U(x, x')=\Delta_{\bar z}^{1/2}(x,x')$ 
and it obeys the following transport equation~\cite{Poisson,WardellOtt}:
\begin{equation} \label{transport VV}
\frac{d\Delta_{\bar z}^{1/2}}{d\bar\lambda}=
\frac{ \left(4-\sigma ^{\alpha }{}_{\alpha }\right)}{2 \bar\lambda }\Delta_{\bar z}^{1/2}
\end{equation}
where $\sigma^{\alpha}{}_{\beta}\equiv \nabla_{\beta}\nabla^{\alpha}\sigma$.
We obtain the van Vleck determinant along the null geodesic $\bar z(\bar\lambda)$ connecting $x'$ and $x$ in Kerr space-time by solving a system of transport equations: Eq.(\ref{transport VV})
together with transport equations for $\sigma^{\alpha}{}_{\beta}$  -- we refer the reader 
to~\cite{WardellOtt} for details.
In principle, we expect the van Vleck determinant $\Delta_{\bar z}(x,x')$ to diverge when $x$ and $x'$ are connected by more than one null geodesic. In a spherically-symmetric space-time
such $x'$ points correspond to points which lie along $\phi=0$ or $\pi$, and  $\Delta_{\bar z}(x, x')$ diverges there~\cite{Casals2009a,Casals:2012px,WardellOtt}; in Kerr space-time the distribution of such points is a lot richer -- see Sec.\ref{sec4}. 
In any case, we calculate  $\Delta_{\bar z}(x, x')$  only along a `direct' null geodesic, which does not re-intersect any other null geodesic that started off at the same initial space-time point $x'$. Therefore, we find that  $U(x, x')=\Delta_{\bar z}^{1/2}(x,x')$ does not diverge anywhere where we need to evaluate it for the results in this paper.

Finally, the biscalar $V(x, x')$ may be expressed (in a normal neighbourhood)  in the following form (e.g.,~\cite{Casals2009})
\begin{equation} \label{eq:V}
V(x, x')=\sum^\infty_{k=0}v_k({ \bf x}, {\bf x'}) \left(t-t'-T_0({\bf x}, {\bf x'})\right)^k\,.
\end{equation}
where the coefficients $v_k$ are regular biscalar functions.
Under a Fourier transformation 
and after taking the high-frequency limit ($M\omega\gg 1$), any terms with $k>0$ will yield contributions of order $O(1/(\omega M) )$, and so the main effect of $V(x,x')$ under this limit is given by $v_0$.

 %---------------------------------------------------------------------------------------------

\subsection{Later images: spectroscopy decomposition}\label{sec:later images}
 
 After the primary signal arrives at the observer, subsequent signals will continue arriving due to the fact that
 part of the null wavefront emitted by the source orbits around the black hole an unlimited number of times.
 These subsequent signals, however, cannot be described by the Hadamard form Eq.(\ref{eqhada}) since they do not lie in a normal 
 neighborhood of the emission point.
 The technique that we will use in order to calculate an approximation to the Green function for points corresponding to the subsequent images is a 
 spectroscopy decomposition in the complex frequency plane.

 When deforming the Fourier integral in the complex frequency plane, the retarded Green function for a field propagating on a black hole background 
 space-time (see, e.g., \cite{Leaver:1986,Ching, Casals2013} in Schwarzschild and  \cite{Yang2013} in Kerr) can be decomposed into three parts: 
 \begin{equation}\label{eqgreenexpansion}
 G_{\rm ret} = G_{\rm HF}+G_{\rm QNM}+G_{\rm BC}\,.
 \end{equation}

 The first part in Eq.(\ref{eqgreenexpansion}), $G_{\rm HF}$, corresponds to an integral along a high-frequency arc. It is believed~\cite{Ching,Casals2013} to describe
  the direct pulse that travels near the future light-cone
and to vanish after a certain finite time corresponding to a point not lying beyond the boundary of the normal  neighborhood.
Since we will only use the spectral decomposition (\ref{eqgreenexpansion}) for points lying outside the normal neighbourhood, we can ignore the contribution
$G_{\rm HF}$.
 
 The second part in Eq.(\ref{eqgreenexpansion}), $G_{\rm QNM}$, corresponds to a sum over the residues at the poles (the so-called quasinormal mode frequencies) of the 
Fourier modes of the Green function. 
This ``quasinormal mode (QNM) part" is related to wave scattering in the strong-gravity region.
In particular, QNMs in the limit of high-oscillation-frequency (i.e., large real part of the QNM frequency or, equivalently, $L\equiv \ell+1/2 \gg 1$, where $\ell$ is the
index for the polar angle eigenvalue in Boyer-Lindquist coordinates)
 in Schwarzschild and Kerr space-times are intimately connected with null geodesics on the photon spheres (e.g., \cite{Mashhoon,Yang2012a}), and there is a one-to-one mapping that relates conserved quantities of these null geodesics to the QNM's frequency, angular eigenvalue, azimuthal quantum number, etc. 
Furthermore, it has been shown  \cite{Dolan2011,Casals2013,Yang2013}  that the QNM part yields the singularity of the Green function $G_{\rm ret}(x,x')$ 
whenever the two space-time points $x$ and $x'$ are connected via a null geodesic.
The null wavefront emitted at $x'$ will wrap itself around the black hole an unlimited amount of times.
Each time that the wavefront orbits around the black hole one more time, a null geodesic which is part of the wavefront will travel outwards to reach the spatial point $\bf x$.
At each instant of time when a null geodesic reaches $\bf x$ the retarded Green function  $G_{\rm ret}(x,x')$  diverges.
Following this correspondence, 
it is clear that it is the  high-oscillation-frequency limit of QNMs that we need to evaluate in order to obtain the null wavefront for the secondary images
(in fact, we expect the high-oscillation-frequency approximation for the QNM contribution to work better as the real radiation frequency $\omega$ becomes larger, which is the physical
limit of interest in this paper).
In Kerr space-time the ``QNM contribution"   is derived in \cite{Yang2013} in the high-oscillation-frequency limit and it is expressed as an
asymptotic expansion
\begin{equation} \label{QNM asympts}
G_{\rm QNM}(x,x') \sim \sum^\infty_{k=1} G_k(x,x')\,.
\end{equation}
The time-domain functions $G_k(x,x')$ are given by
 \begin{equation}
\label{eqifact3}
G_{k} = \left\{
\begin{array}{cl}
\displaystyle (-1)^{k/2} \pi \, \chi_{\rm e} \, \delta[\pi k+g_{\rm e}(\mu_{\rm e})],&k\ {\rm even}\,, \\
\\ 
\displaystyle (-1)^{(k+1)/2} \frac{\chi_{\rm o} \, H [-\pi k-g_{\rm e}(\mu_{\rm e})]}{-\pi(k+1)-g_{\rm o}(\mu_{\rm o})}\,, &k\ {\rm odd}\,,
\end{array}\right.
\end{equation}
where $\mu_{\rm e/o}$ characterize the ``shape" of a geodesic connecting $x$ and $x'$, $g_{\rm e/o}$ are the phase functions and $\chi_{\rm e/o}$ are the excitation amplitudes--- they are all regular functions of $x$, $x'$ and the parameters of the black hole
 (their physical meanings and detailed expressions are given in Sec.V of \cite{Yang2013}).
 The asymptotic expansion of Eq.(\ref{QNM asympts}) is valid in the limit when the space-time points $x$ and $x'$ are joined by a null geodesic.
Here, $k$ is a positive integer index that labels the  singularities  of the Green function at different times as the wavefront orbits around the black hole.
In Schwarzschild space-time, $k$ is also a label for the amount of times that the null geodesic joining $x$ and $x'$ has crossed a caustic
and we expect this to still be true 
in Kerr space-time. We note that here and throughout this subsection by `caustic' we mean a space-time point (say, $x$) where two or more null
geodesics that started off at the same space-time point (say, $x'$) meet.
This meaning is different from that generally used in optics, which we define in Sec.\ref{sec4} and use in the rest of the paper, where the optical phase is required to be stationary .
We shall therefore refer to the latter as an `optical caustic' to differentiate it from the `caustic' that we have just defined and that we use in this subsection.

  By applying a Fourier transform to the time-domain functions $G_k$ we find that the even-$k$ modes scale as $O(1)$ in the frequency domain, whereas the odd-$k$ modes scale as $O(1/(\omega M))$ for large-frequency $\omega M$.
 As mentioned in the introduction, we assume the wavelength of the radiation to be much smaller than the size of the black hole and, therefore, in this $\omega \gg 1/M$ limit
 we neglect the odd-$k$ modes.
The total contribution from the even-$k$ modes in the frequency domain is just\footnote{After the Fourier transformation, the 
$t$-dependence of $\mu_e$ and $\chi_e$ is replaced by the specific value of 
 $t$
 such that the corresponding
 $\delta$-Dirac distribution in Eq.(\ref{eqifact3}) becomes singular. Therefore,  in the frequency domain they are functions of $k$, $\bf x$ and $\bf x'$.}:
\begin{align}\label{eqqnmg}
G_{\rm QNM,e}^{\omega} &=\sum^\infty_{{\rm even}\ k=2}(-1)^{k/2} \frac{\pi\, \chi_{\rm e}}{\Omega_R(\mu_{\rm e})} e^{i \omega T^{\rm e}_{k}(\bf x,\bf x')}\nonumber \\
&=\sum^\infty_{{\rm even}\ k=2}(-1)^{k/2}A_k^{\rm e}({\bf x},{\bf x'}) e^{i \omega T^{\rm e}_{k}(\bf x,\bf x')}\,.
\end{align}
where $\Omega_R=\Omega_R(\mu_{\rm e})$ is the real part of the QNM frequency (in the large-$L$ limit) divided by $L$~\cite{Yang2012a}, $A^e_k \equiv \pi \chi_e/\Omega_R(\mu_e)$ and 
$T^e_k$ is  the value of the Boyer-Lindquist time such that $g_e(\mu_e)=-\pi k$.
That is, $T^e_k$ corresponds to an approximation of the time at which the null geodesic that starts at $x'$  and passes through $k$ caustics reaches the point $x$.

The third and last~\footnote{We note that in Kerr space-time there may be extra branch cuts in the complex-frequency plane due to the spheroidal eigenfunctions and eigenvalue
but, in that case, we also expect their Fourier transform to be of order $o(1)$ for large frequency.}  
of the contributions to the Green function in Eq.(\ref{eqgreenexpansion}), $G_{\rm BC}$, 
corresponds to  the branch cut of the modes $G^{\omega}_{\rm ret}(\bf x,\bf x' )$ starting at the origin in the complex-frequency plane (e.g.,~\cite{Leaver:1986,Ching,Casals2012a}).
This contribution is related to wave scattering by the Coulumb-type potential and it yields the well-known power-law tail decay at late-times~\cite{Price:1972,Leaver:1986,Casals2012a,Hod}. 
This part of the Green function is smooth in the time domain
at  least after the time when the first, `direct' null geodesic has joined $x$ and $x'$~\cite{Casals2012c,Casals2013,Casals2012b,Casals2012a}.
Furthermore, any divergence of $G_{\rm BC}$ before the arrival of the `direct' null geodesic is purely due to the 
exponential Fourier factor
and so the contribution in the frequency domain from the Fourier transform of  $G_{\rm BC}$ is 
of little-oh order $o(1)$~\cite{Casals2012c}.
As a result, 
 we neglect this piece of the Green function in the high-frequency limit.

%-----------------------------------------------------------------------------------------------------------------------------------------------------------------------------------------------------------

  \section{Evaluating the wave propagation and POAM}\label{sec:wave}

The analysis in the previous section
 implicitly assumes that the wave emission is coherent within any time interval we are considering, as we have taken the integration-lower-bound to be `$-\infty$' in Eq.~(\ref{eqconv}). This assumption may not be astrophysically realistic, because the typical radio telescope's bandwidth is of the order of GHz, and the time difference between the direct pulse and all the later pulses linearly scales with the black hole mass $M$, which is of the order of micro seconds if $M \sim M_{\odot}$ (solar mass). In particular, for supermassive back holes with masses $\ge 10^5 M_{\odot}$, the coherence time of the wave is much shorter compared to $M$, in which case the contributions
from the $\delta$-Dirac distributions in the $G_{\rm QNM}$ are no longer important. 
Nevertheless, for the sake of completeness, in the following analysis we shall consider two limiting cases: the coherence length of the wave being much longer or shorter than the black hole size. We shall find that the  POAM spectra that a distant observer receives are completely different in these two scenarios.

\subsection{``Infinite" coherence length}\label{sec3a}

With the coherence length much longer than the black hole size, the coherent part of the wave that arrives at an observer can be described by the sum of terms in Eq.~(\ref{eqdirg}) and Eq.~(\ref{eqqnmg}). These terms all contain fast-oscillating phase factors which are proportional to $\omega$, and slowly-varying amplitude factors.
From now on we will consider the observer to be located far away from the black hole and the source of radiation.
The observation plane is orthogonal to the direction 
of propagation of the wave, i.e., constant-$r$ surface (locally a plane),  $r=\robs$.
For a fixed emitter location $\bf x'$, the phase factors $T_0$ and $T^{\rm e}_k$ are just functions of 
the observer location 
${\bf x}=(r_\indobs,\theta_\indobs,\phi_\indobs)$, in Boyer-Lindquist coordinates $(r,\theta,\phi)$.
We choose a point with coordinates ${\bf x_0}=(\robs,\theta_0,\phi_0)$ on the observation plane as the origin, and expand the phase functions as
\begin{equation} \label{eq:exp T}
T({\bf x})\approx T({\bf x_0})+\left. \frac{\partial T}{\partial \theta_\indobs}\right|_{\bf x_0} \delta \theta_\indobs+\left.\frac{\partial T}{\partial \phi_\indobs}\right|_{\bf x_0}  \delta \phi_\indobs \,,
\end{equation}
where $\bf x$ is a point  on the observation plane,
 $T({\bf x})$ generically denotes $T_0$ and $T^e_k$, and from now on we omit $\bf x'$ from the argument of $T(\bf x)$ and from that of the amplitudes
 (it is understood in Eq.(\ref{eq:exp T}) and similar expressions below that $\bf x'$ and $\robs$ are fixed).

For later convenience, we further define a new coordinate system on the observation plane to be $(x_1, x_2) \equiv ( \delta \theta_\indobs,  \sin\theta_0 \delta \phi_\indobs)$ and
we define $\obscoord\equiv x_1+i x_2$, with its complex conjugate given by $\obscoord^*=x_1-i x_2$ and its absolute value by $|\obscoord| = \sqrt{x^2_1+x^2_2}$.
The latter is related to the cylindrical radial coordinate introduced in Sec.\ref{sec2} as $\rho=|\obscoord|\robs$.
The expansion of the phase in terms of the new coordinates is just
\begin{align}\label{eqlinT}
T({\bf x}) \approx& 
T({\bf x_0})+\left.\frac{\partial T}{\partial \theta_\indobs}\right|_{\bf x_0} x_1+\left.\frac{\partial T}{\partial \phi_\indobs}\right|_{\bf x_0} \frac{x_2}{ \sin \theta_0} \nonumber \\
=&T({\bf x_0})+\left.\frac{\partial T}{\partial x_1}\right|_{\bf x_0}x_1+\left.\frac{\partial T}{\partial x_2}\right|_{\bf x_0} x_2\,.
\end{align}

As the detector is usually far away from the source, the characteristic values of $x_{1,2}$ or $|\obscoord|$ are of the scale of $b/\robs$ where,
as introduced earlier, $b$ is the size of the detector. In addition to the ``high frequency" assumption $\omega M \gg 1$, we further assume that 
\begin{equation}
\eta \equiv \omega M b/\robs \ll1\,,
\end{equation}
 which is equivalent to saying that the instrument size $b$ is much smaller than the Airy disk size $\lambda \robs /M \sim 2\pi \robs/(\omega M)  $ of the diffraction pattern. For a given radiation frequency $\omega$, in order to maximize $\eta$, one can either build longer telescope arrays or target the black holes with the largest angular sizes $M/\robs$ in the sky. The best known supermassive black-hole candidates would be the one in the 
galaxy Messier $87$ or Sagittarius A* in our own galaxy. Take,  for example, the millimeter-wavelength sources near Sagittarius A* if the size of the telescope array is comparable to the Earth radius, i.e. $b \sim 6\times 10^3$km, the resulting $\eta$ is about $0.6$.

Under the above assumptions, we can expand the phase factors as
\begin{align}\label{eqlinexp}
e^{i \omega T({\bf x})} \approx & \,e^{i \omega T({\bf x_0})} \times \nonumber \\
&\left [1+i \omega \left (\left.\frac{\partial T}{\partial x_1}\right|_{\bf x_0}x_1+\left.\frac{\partial T}{\partial x_2}\right|_{\bf x_0} x_2 \right )\right ].
\end{align}

It is straightforward to show that Eq.~(\ref{eqlinexp}) only generates $l=\pm 1$ components in the POAM spectrum (it also generates a $l=0$ component but it
has zero orbital angular momentum), because of the linear dependence in $x_{1,2}$. If we included higher order terms in $x_{1,2}$ in the power series expansion in Eq.~(\ref{eqlinT}), the $|l| \ge 2$ components would also show up in the POAM spectrum, but with weaker spectra weights since $\sqrt{w_l/w_1} \propto \eta^{|l|-1} \ll 1$, $\forall |l|\ge 2$. Therefore, in the following analysis we focus on the $l=\pm 1$ components of the spectrum.

Combining Eqs.~(\ref{eqdirg}), (\ref{eqqnmg}), (\ref{eqandec}) and (\ref{eqpsiGreen}) and Eq.(\ref{eqlinexp})
 in Eq.~(\ref{eqpsiGreen}),
we obtain the following expression for $\psiFl{\pm 1}$: 
\begin{align}\label{eqpsi1coh}
&\psiFl{1} =\frac{iN|\obscoord| \,\omega}{2} A_0({\bf x_0})e^{i \omega T_0({\bf x_0})}\left .\frac{\partial T_0}{\partial \obscoord} \right |_{\bf x_0}\nonumber \\
&+\frac{iN|\obscoord| \,\omega}{2}\sum^\infty_{{\rm even}\ k=2}(-1)^{k/2}A_k^{\rm e}({\bf x_0}) e^{i \omega T^{\rm e}_{k}({\bf x_0})}\left .\frac{\partial T^{\rm e}_k}{\partial \obscoord} \right |_{\bf x_0}\,,\\
&\psiFl{-1} =  \,\psiFl{1}(\obscoord \rightarrow \obscoord^*)\,,
\end{align}
where $A_0({\bf x_0})$ and $A_k^{\rm e}({\bf x_0})$ are approximations to $A_0({\bf x})$ and $A_k^{\rm e}({\bf x})$, respectively (omitting the argument $\bf x'$).
For any real-valued function $T(x_1,x_2)=T(\obscoord,\obscoord^*)$, we notice that the following relation always holds:
\begin{equation}\label{eqpTid}
\left ( \frac{\partial T}{\partial \obscoord}\right )^* = \frac{\partial T}{\partial  \obscoord^*}\,.
\end{equation}

\begin{figure}[t,b]\centering
\includegraphics[width=0.9\columnwidth]{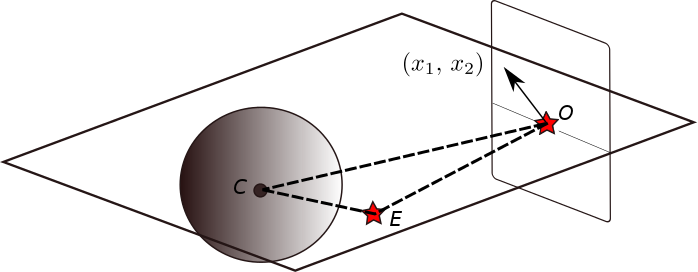}
\caption{(Color online.) 
An illustration of the relative positions of the emitter $E$ (at $\bf x'$), the origin of the observation plane $O$ (at $\bf x_0$) and a Schwarzschild black hole centered at the point $C$. 
The arrow points to a point with coordinates $(x_1, x_2)$ on the observation plane. The triangle that contains the points $E, O, C$ lies on the mirror-symmetry plane.}
\label{fig:sBH}
\end{figure}

\subsubsection{Schwarzschild black hole}

In the case that the host black hole is a Schwarzschild black hole, the phase functions $T_0$ and $T^{\rm e}_k$ are symmetric with respect to the plane spanned by the emitter, the center of the black hole and the receiver. Due to this mirror symmetry, the function $T({\bf x})$ in Eq.~(\ref{eqlinT}) only depends on one variable, which is the projection of $(x_1,x_2)$ onto the mirror-symmetry plane -- see Fig.~\ref{fig:sBH}. This one-parameter dependence directly leads to the finding that all the complex numbers $\partial T/\partial \obscoord$ in Eq.~(\ref{eqpsi1coh}) share the same argument in the complex plane, i.e., the ratio between any two of these numbers is real. Combining this fact with Eq.~(\ref{eqpTid}), it is straightforward to show that $|\psiFl{1}|=|\psiFl{-1}|$, and similarly for the POAM spectrum weights with $|l| \ge1$. In other words, Schwarzschild POAM spectra are always symmetric.

\subsubsection{Kerr black hole}

It is however reasonable to expect asymmetric POAM spectra for radiation sourced near generic Kerr black holes, as originating from the interference between the direct signal and the secondary signals. In the geometrical optics picture, in order to generate non-zero POAM along the optical axis, the propagation directions of these rays must not lie on the same plane
\footnote{The propagation direction of these rays is not along the $r$ axis, because the phase functions $T(x_1,x_2)$ of these rays are not constant on the celestial sphere.} 
 (this condition is not satisfied in the case of a Schwarzschild black hole) and the rays with constant phase have to offset each other at the places where a detection is made. The first condition is satisfied if $\partial T_0/\partial \obscoord$ and $\partial T^e_k/\partial \obscoord$ do not have the same argument in the complex plane, and the second condition is true when $\omega T_0({\bf x_0})$ and $\omega T^e_k({\bf x_0})$ are not in-phase. It is generic to meet both conditions for emitters near a rotating black hole.

 \begin{figure}[t,b]\centering
\includegraphics[width=0.9\columnwidth]{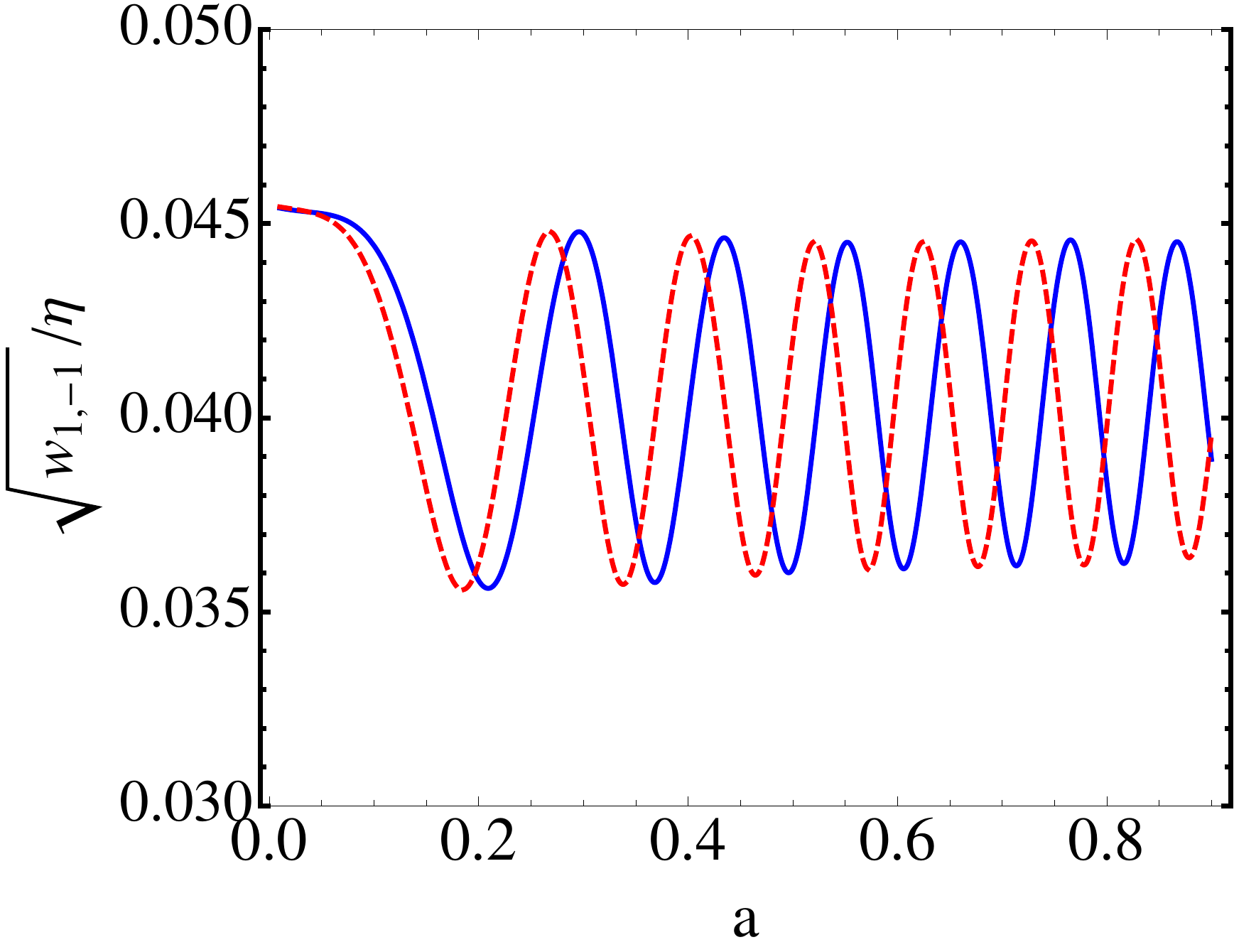}
\includegraphics[width=0.85\columnwidth]{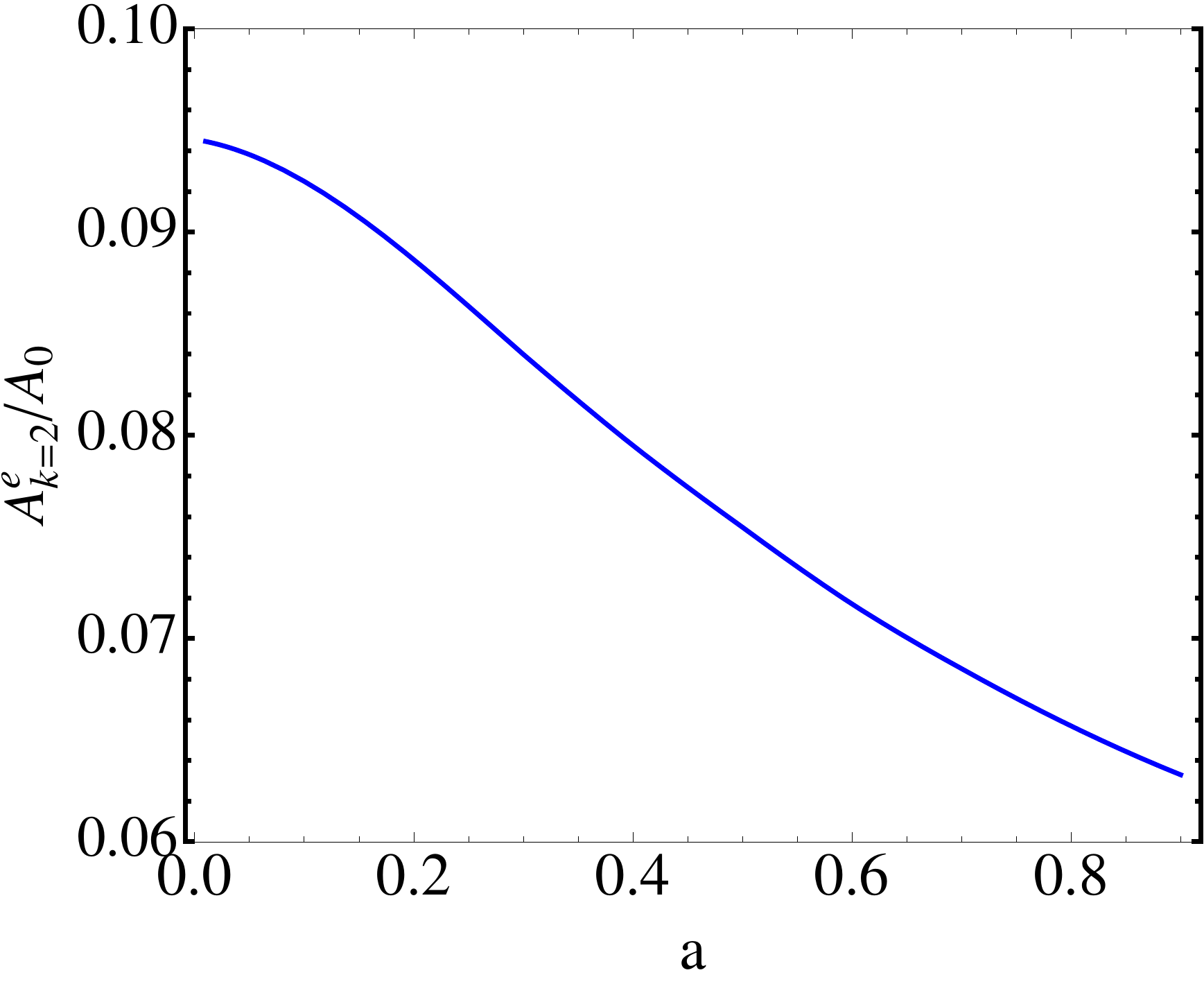}
\caption{(Color online.) Top figure: $\sqrt{w_1}/\eta$ (blue solid line) and $\sqrt{w_{-1}}/\eta$ (red dashed line) for different black hole spins.
The emitter and the receiver are respectively located at $(r_\indem,\theta_\indem,\phi_\indem)=(8M, \pi/2, 0)$ and 
$(r_\indobs,\theta_\indobs,\phi_\indobs)=(10^4 M, 2\pi/3, 0)$.
 For illustration purposes, $\omega$ is chosen to be $4\pi/M$ to avoid too many oscillations. The spectral asymmetry vanishes for Schwarzschild black hole ($a=0$), and oscillates as varying black hole spin, because of the changing phase difference between the primary image and the secondary image $k=2$. Bottom figure: the ratio between $A^{\rm e}_{k=2}$ and $A_{\rm 0}$ for the same parameter settings.}
\label{fig:wpm1}
\end{figure}

Since, on dimensional grounds, the phase functions $T_0$ and  $T^{\rm e}_k$ are proportional to M, we find that $\psiFl{\pm 1} \sim |\obscoord| \omega M$ 
and $\sqrt{w_{\pm 1}} \sim \omega M b/\robs=\eta$ for small $\eta$. In order to illustrate this asymmetry, we choose the emitter's and receiver's locations to be, respectively, ${\bf x'}=(r_\indem,\theta_\indem,\phi_\indem)=(8M, \pi/2, 0)$ and 
${\bf x}=(r_\indobs,\theta_\indobs,\phi_\indobs)=(10^4 M, 2\pi/3, 0)$, the radiation frequency to be $\omega=4\pi/M$, and,
using the techniques described in Sec.\ref{sec:GF} we calculate $A^{\rm e}_k$, $A_0$, $\psiFl{\pm 1}$ and $w_{\pm 1}$.
In the top Fig.~\ref{fig:wpm1} we plot $\sqrt{w_{\pm 1}}/\eta$ with varying black hole spin $a$.
This plot
shows that the spectral asymmetry $|\sqrt{w_1}-\sqrt{w_{-1}}|$ is of the order of $10^{-2}\eta$. In making this plot, we only kept  the primary pulse and the secondary pulse with $k=2$ (note $g_{\rm e}(\mu_{\rm e})=-\pi k$ does not have a solution for $k=0$ in this case, and the summation in Eq.~(\ref{eqpsi1coh}) starts from $k=2$). 
In the  bottom Fig.\ref{fig:wpm1} we plot $A^{\rm e}_{k=2}/A_0$ with varying black hole spin $a$.
This plot shows that the secondary images can be almost as bright as the primary image: $A^{\rm e}_k/A_0 = O(0.1)$, 
depending on the emitter's and receiver's locations, as well as the spin of the black hole.

Although it is unnatural to expect the coherence length of the wave to be longer than the size of astrophysical black holes (a coherence length which is comparable to the size of a solar-mass black hole requires the detection bandwidth to be below MHz; for supermassive black-holes, such as Sagittarius A*, the corresponding bandwidth is below $0.1$ Hz), 
the primary and secondary signals from the same-time emission are still coherent with each other, and they arrive at Earth at different times.
This means that there is nonzero correlation within the time-series data of the field $ \langle E(t)E(t+\tau') \rangle$ (where  $E$ is the electromagnetic field and we here neglect its polarization), if we set the delay-time $\tau'$ to be the time-lag between the direct signal and other secondary signals \cite{Boyle}. The significance of this correlation should be comparable to $A^{\rm e}_k/A_0$, which could be at least of  order  $O(0.1)$.

\subsection{Incoherent emission}\label{sec3b}

In this subsection, we consider a more realistic scenario where the coherence length of the emission is much shorter than the size of the black hole. In this case, the signal has a contribution from the primary pulse that travels on the light cone and from the coherent follow-ups
(which here they correspond to the tail $V(x,x')$, not to the secondary pulses)
 which lag only a short amount of time (compared to the characteristic time-lag between different pulses) after the primary pulse.

Because of the finite-coherence effect, we shall only consider the contribution from the primary pulse and the follow-up signal described by $V(x, x')$.
Physically, this is justified by noticing that the part of the follow-up signal which lags  far behind the primary signal contributes insignificantly to the total coherent wave. 
 Mathematically,
this may be achieved, for example, by assigning a small imaginary part to the wave frequency, i.e. replacing $\omega$ by $\omega +i \epsilon$ with $\epsilon>0$, and,
at  the end of the calculation, taking the limit $\epsilon\to 0^+$. As a result of this operation, the coherent wave is evaluated to be:
\begin{align} \label{psi incoherent}
\psiF(x) \approx  N \left [A_0({\bf x}, {\bf x'}) 
+\frac{v_0({\bf x},{\bf x'})}{i \omega}\right ] \,e^{-i \,\omega T_0}\,.
\end{align}
In order to fully incorporate the finite-coherence effect, the monochromatic emission from the source should
 be generalized to  one with a frequency spectrum. In this case, the wave arriving at the distant observer would be
\begin{equation}\label{eqmulfreq}
\Psi(x) = \int d \omega f(\omega) \psiF ({\bf x})e^{-i\omega t}\,,
\end{equation}
where $\langle f(\omega) f^*(\omega') \rangle = 1/(2 \pi) S_{\omega}\delta(\omega-\omega')$ ($\langle \rangle$ stands for the ensemble average in random process) and $S_{\omega}$ is the normalized frequency-spectra density
of the source: $\int d \omega S_\omega=1$. In the following analysis, however, we shall stick to the monochromatic approximation in discussing the effect of interference --- it is straightforward to include multiple frequencies using Eq.~(\ref{eqmulfreq}), and it turns out that the monochromatic case already captures the main physics (about POAM spectrum) here.

The spectra components $\psiFl{0}$ (which plays a part via the overall normalization $I$ in Eq.(\ref{eqnorm})) and  $\psiFl{\pm 1}$ can be obtained by performing the angular integration described in Eq.~(\ref{eqandec}), thus yielding
\begin{align}
\psiFl{0} \approx& N \left [A_0({\bf x_0}) -\frac{ v_0}{i\omega}\right ]e^{i \omega T_0} \,,\nonumber \\
\psiFl{+1}\approx&\frac{iN |\obscoord| \,\omega}{2} \left [A_0 -\frac{ v_0}{i\omega }\right ]e^{i \omega T_0}\left.\frac{\partial T_0}{\partial \obscoord}\right|_{\bf x_0} \nonumber \\
&+\frac{N|\obscoord| }{2}e^{i \omega T_0}\left.\frac{\partial A_0}{\partial \obscoord}\right|_{\bf x_0}\,,\label{eqfpsip}\\
\psiFl{-1} \approx & \,\psiFl{1}(\obscoord \rightarrow \obscoord^*)\,.\label{eqfpsim}
\end{align}
The above expressions show that  in the large-frequency limit these tail effects are subdominant to those from the direct wavefront.
That is, $\psiFl{\pm 1}$ are still dominated in the  high-frequency limit by the direct pulse that travels on the light-cone, with a
magnitude of order $|\obscoord| \omega M$. However, the term
with $\partial T_0/\partial \obscoord$ in
 Eq.~(\ref{eqfpsip}) 
and the corresponding term in Eq.~(\ref{eqfpsim}) contribute evenly to the magnitude of $\psiFl{\pm 1}$ or $w_{\pm 1}$.
Therefore, the POAM spectral asymmetry arises from the beating between the amplitude and phase variations of the direct signal, the latter variation
being suppressed by a $1/{(\omega M)}$ factor. In other words, although the spectrum weight is still $\sqrt{w_{\pm 1}} \sim \eta$, the spectral asymmetry scales as $|\sqrt{w_1}-\sqrt{w_{-1}}| \sim  b/\robs$ for small $\eta$. In fact, this net POAM is expected to originate from 
the frame-dragging effect of rotating black holes, and it vanishes in a Schwarzschild background. As suggested by Eq.~(\ref{eqfpsip}), if the amplitude modulation does not follow the same direction as the phase modulation on the celestial sphere, these ``unbalanced" rays will generate a nonzero contribution to the angular momentum along the $r$-axis. From another point of view, the black-hole rotation breaks the ``mirror symmetry" in detection that we discussed for Schwarzschild black holes, and the beating of phase and amplitude modulations introduces a net POAM, even though it is extremely small.

For millimeter-wavelength sources near Sagittarius A*, even if the size of the telescope array is of the scale of the Earth radius $6 \times 10^3$ km, the spectral asymmetry $b/\robs$ is of the order of $10^{-14}$ which is way too weak to be measured. 
On the other hand, the pure spectrum weights $\sqrt{w_{\pm 1}} \sim \eta$ 
are of the order of $0.6$. Therefore, it seems feasible that  these $l \neq 0$ spectrum components\
could be resolved
 by coherent detection by  spatially-separated telescopes. 
 However, as we recall from the discussion in Sec.~\ref{sec2},  the symmetric part of the POAM spectrum is degenerate with the tilt angle of the observation plane. In order to
extract information about the black hole  from the POAM measurement, it might be necessary to use the asymmetric part of the spectrum, which is extremely small in this case.

We note that the result we have obtained in this subsection of weak POAM spectrum asymmetry  is different from  that  obtained in \cite{Tamburini}.
The reason is that, as mentioned, the physical setttings are different, since in \cite{Tamburini} they 
consider an extended source, as opposed to our
pointlike
 source, while they
assume  the emission from the whole accretion disk around a black hole  to be spatially coherent.
 Thus, their main asymmetry effect comes from the interference between the radiation from different parts of the disk,
 as opposed to it being due to the light-twisting effect that we have studied in this subsection (or to
 interference between different rays emitted from the same point and arriving at different times
 as we saw in the previous subsection).

 %-----------------------------------------------------------------------------------------------------------------------------------------------------------------------------------------------------------

\section{Receivers near an Optical Caustic}\label{sec4}

The analysis in Sec.~\ref{sec:wave} is invalid if the receiver is located close to an optical caustic (defined below).
At such a point,  the 
Hadamard form Eq.(\ref{eqhada}) for the Green function 
and
the
high-oscillation-frequency asymptotics of Eq.(\ref{QNM asympts}) for the QNM contribution to the Green function 
 both fail. 
As we shall show below, the wave intensity is  amplified near an optical caustic, which increases the detectability of the source. 
 While for a pointlike source 
\footnote{It could be star, or a flare with extraordinary brightness in the accretion disk.}
near a black hole we expect  the chance of Earth being located near an optical caustic 
point of its radiation to be  small, we expect the chances to be significantly higher if the source is {\it extended}.
It  still remains physically interesting to understand the wave near an optical caustic and to exploit the corresponding POAM spectrum. 

In the following analysis, we shall heavily apply the techniques and results of {\it catastrophe optics}, which is  expounded in the excellent text by Berry and Upstill \cite{berry}. Interested readers can also find detailed numerical investigations of the optical caustic structure in Kerr space-time in \cite{Bozza}, and its relation with X-ray variability in \cite{Rauch}.
We  now briefly introduce the main concepts and terminology.

\begin{figure}[t,b]\centering
\includegraphics[width=1.0\columnwidth]{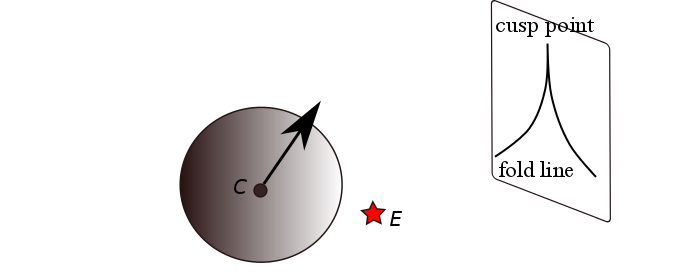}
\caption{(Color online.) An illustration of potential optical caustic structure on the celestial sphere/observation plane, due to an emission source E near a spinning black hole centered at C. {\it Fold} lines meet each other at a {\it cusp} point.
}
\label{fig:cuspfold}
\end{figure}

At a given distance, the receiver's sky-location can be characterized by the two-parameter family $(\theta_f, \phi_f)$ or $(x_1, x_2)$, which are often referred
to as {\it control parameters} in {catastrophe optics} and which we denote by $C_i$, $\forall i$. In its turn, the  angles of the emitted rays are usually referred to as {\it state variables}
and we denote them generically by $s_j$, $\forall j$. 
The `action function' $T_0(\bf x, \bf x')$ (see Eq.(\ref{eqdirg})), when viewed as a function of the null rays (for fixed $\bf x$ and $\bf x'$) is generally a multi-valued function.  
It can be made into a single-valued function by including a functionality on the state variables
 (which allow to uniquely characterize a null ray).
Such single-valued function is called  a ``generating function"  and we denote it by $\Phi(s_j; C_i)$.
Null rays are those paths that extremize the generating function, i.e., such that $\partial \Phi/\partial s_j=0$, $\forall j$.
Optical caustics, then, correspond to  {\it singularities} of this gradient map.
 A {\it structurally stable singularity} is a singularity of the above gradient map such that if it is perturbed, it is related to the new singularity by a diffeomorphism of $C_i$.
For example, the singularities or optical caustics in Schwarzschild space-time are structurally unstable, as they are susceptible to small perturbations of the space-time. 
 However, a {\it fold} line or a {\it cusp} point (described below and illustrated in Fig.~\ref{fig:cuspfold}) in Kerr space-time are generically stable singularities.
As explained in \cite{berry}, all the structurally stable singularities
can be classified into different equivalence classes, called  {\it catastrophes},  with a generating function associated to each class. 

\subsection{Cusp point}
\begin{figure}[t,b]\centering
\includegraphics[width=1.0\columnwidth]{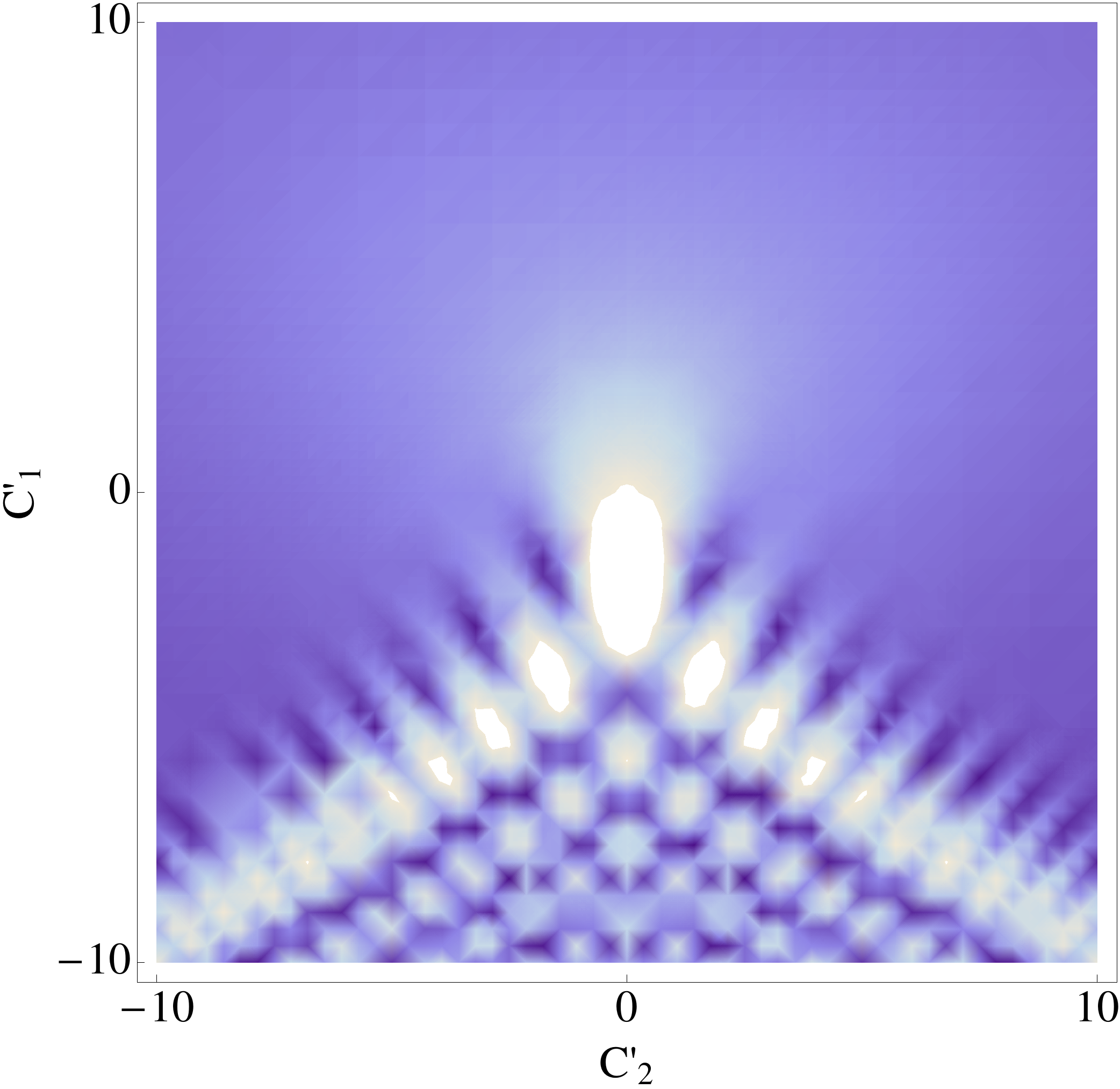}
\caption{(Color online.) A contour plot of the 
integral in Eq.(\ref{eqpsic2}) (which is proportional to the wave function near a cusp point), 
where a darker region corresponds to a region with a higher wave amplitude and
where the axis corrrespond to generic rescaled control parameters  $C'_{1,2}$. 
}
\label{fig:cusp}
\end{figure}

It turns out that the only  catastrophe with a $2$-dimensional control-parameter space has the following generating function:
\begin{equation}
\Phi(s; C_1, C_2) = \frac{s^4}{4}+\frac{s^2}{2}C_2+s C_1\,,
\end{equation}
where $s$ refers to the state variable that matters for the singularity classes
(the other state variable does not affect the singularity class)
 and $C_{1,2}$ are the control variables. In the case we are considering, $C_{1,2}$ can be mapped to $x_{1,2}$ under proper rotation and rescaling (here, $C_{1,2}$ and $x_{1,2}$ are all dimensionless, so the rescaling factors should also be dimensionless). The singularity point at $C_{1,2}=0$ is usually called a ``cusp" in catastrophe theory. It is also worth noting that one can always add any additional independent state variable $s'$, with associated generating function that are at most quadratic (for example ${s'}^2$), and it does not affect the singularity structure. With this generating function,  the wave function 
\footnote{Strictly speaking, $\psiF$ in Eq.(\ref{eq:wave cusp})
 should be referred to as the spatial part of the wave function, but given the simple relation Eq.(\ref{eqpsiGreen}), in this section
  we will loosely refer to it as the wave function itself.}
 near a cusp is approximately:
\begin{equation}\label{eq:wave cusp}
\psiF(C_1, C_2) \propto \sqrt{\omega} \int_{-\infty}^{\infty} ds \,e^{i \omega \Phi(s; C_1, C_2)}\,.
\end{equation}
The proportionality factor in Eq.(\ref{eq:wave cusp}) is independent of $\omega$;
 we will not write out such proportionality factor throughout this section.
See Fig.~\ref{fig:cusp} for an illustration of the wave function near a cusp point.
In order to extract the frequency dependence of $\psiF$, we make a transformation of the variables $s$ and $C_{1,2}$:
\begin{equation}\label{eqpsic1}
s' \equiv \omega^{1/4}s,\, C'_1\equiv \omega^{3/4}C_1,\, C'_2\equiv\omega^{1/2}C_2\,.
\end{equation}
Then $\psiF$ becomes
\begin{equation}\label{eqpsic2}
\psiF(C_1,C_2)\propto \omega^{1/4} \int_{-\infty}^{\infty} ds' e^{i \Phi(s'; C'_1, C'_2)}\,,
\end{equation}
which implies that the amplitude of the wave is amplified by a $(\omega M)^{1/4}$ factor near a cusp. For millimeter-wavelength sources  near a supermassive black hole with mass comparable to Sagittarius A*, this is an order of $10^7$ amplification in the intensity.

We can also determine the POAM spectrum from Eq.~(\ref{eq:wave cusp}) or Eq.~(\ref{eqpsic2}). Let us 
denote by $(C^0_1, \,C^0_2)$ the control variables corresponding to the location of the origin of the telescope array.
We then expand the control variables as $C_1=C^0_1+\delta C_1,\, C_2=C^0_2+\delta C_2$. 
Correspondingly, the wave function on the observation plane can be expanded as
\begin{equation}\label{eqcuspexp}
\psiF \propto \omega^{1/4} \int_{-\infty}^{\infty} d s' \left (1+ i\frac{\partial \Phi}{\partial C'_1}\delta C'_1+i\frac{\partial \Phi}{\partial C'_2}\delta C'_2\right)e^{i \Phi_0}\,,
\end{equation}
where we have defined
$\Phi_0 \equiv \Phi(s'; {C'}^0_1, {C'}^0_2)$.

Performing the POAM decomposition with respect to the above wave function, we find that the main contribution comes from the term with $\delta C'_1$  in Eq.~(\ref{eqcuspexp}), which  
gives $\sqrt{w_{\pm 1}} \sim (\omega M)^{3/4}b/\robs$. The interference between the terms with $\delta C'_2$ and with $\delta C'_1$ generates the asymmetry in the POAM spectrum, which is $|\sqrt{w_1}-\sqrt{w_{-1}}| \sim (\omega M)^{1/2} b/\robs$. For a millimeter-wavelength source near Sagittarius A*, if Earth happens to be close to a cusp point in the celestial sphere, the inferred POAM spectral asymmetry is about $10^{-7}$, which is still too weak to be measured.

\subsection{Fold line}

\begin{figure}[t,b]\centering
\includegraphics[width=1.0\columnwidth]{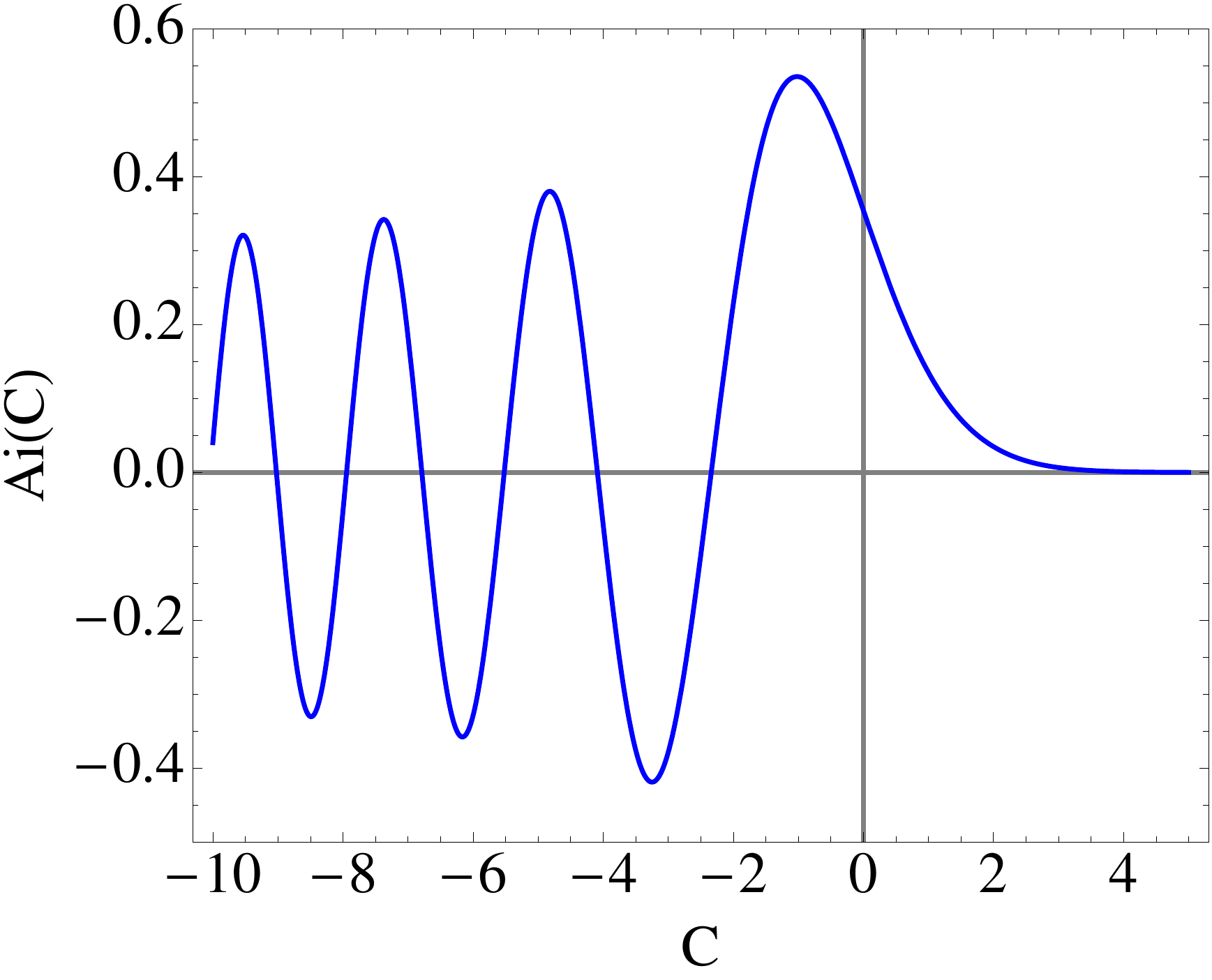}
\caption{(Color online.) 
Plot of the Airy function, which describes the wave function dependence on a control variable $C$ near a fold point via Eq.(\ref{eqpsifoldline}).
}
\label{fig:fold}
\end{figure}

There is another possibility, perhaps with a higher chance than a cusp, which is that Earth is located close to the singularity described by a {\it fold} line. This happens when only one of the two coordinates on the celestial sphere is the ``control variable", and the phase is linear in the other variable. According to catastrophe theory, there is only one catastrophe that has a one-dimensional control space, and its characteristic polynomial is given by
\begin{equation}
\Phi(s; C_1) = \frac{s^3}{3}+s C_1\,,
\end{equation}
with the line associated with $C_1=0$ usually referred to as a ``fold" line. 
Let us denote the coordinate along the fold line as $C_2$ (for a specific value of $C_2$ we have a fold point, and as the value of $C_2$ varies continuosly we have 
a fold line) and the generating function along the fold line as $\Theta(C_2)$. Then the wave function near a fold line is
\begin{align}\label{eqpsifoldline}
\psiF(C_1, C_2) & \propto \sqrt{\omega} \int_{-\infty}^{\infty} ds \,e^{i \omega [\Phi(s; C_1)+\Theta(C_2)]}\nonumber \\
&= \omega ^{1/6}\int_{-\infty}^{\infty} ds' \,e^{i  [\Phi(s'; C'_1)+\omega \Theta(C_2)]} \nonumber \\
&= 2\pi \omega^{1/6} {\rm Ai}(C'_1)e^{i \omega \Theta(C_2)}\,,
\end{align}
where ${\rm Ai}(x)$ denotes the Airy function which is real-valued -- see Fig.~\ref{fig:fold} for an illustration. The new variables $s'$ and $C'_1$ are here defined as
\begin{equation}
s' \equiv \omega ^{1/3}s, \quad C'_1\equiv \omega^{2/3}C_1\,.
\end{equation}

We can see from Eq.~(\ref{eqpsifoldline}) that the wave intensity is amplified by a factor of $(\omega M)^{1/3}$ near a fold line, which is brighter than a generic point on the celestial sphere, but dimmer than a cusp point.
Similar to the analysis performed near a cusp point, we expand the celestial coordinates in the observation plane as: $C_1=C^0_1+\delta C_1,\, C_2=C^0_2+\delta C_2$, and the wave function can also be expanded as
\begin{align}\label{eqfoldexp}
&\psiF(C_1, C_2) -\psiF(C^0_1,C^0_2) \propto \,2\pi \times\\
& \omega^{1/6} \left [\frac{\partial {\rm Ai(C'_1)}}{\partial C'_1}\delta C'_1+i\frac{\partial \Theta}{\partial C_2}\delta C_2 {\rm Ai}(C'_1)\right ]e^{i \omega \Theta}.
\end{align}

Correspondingly, we obtain the strength of $\sqrt{w_{\pm 1}}$ to be proportional to $\eta$, which comes from the phase variation along the fold line;  the asymmetry $|\sqrt{w_1}-\sqrt{w_{-1}}|$ is directly proportional to $(\omega M)^{2/3}b/\robs$ and it is attributed to the phase beating between $\Phi$ and $\Theta$. For a millimeter-wavelength source near Sagittarius A*, if Earth happens to be close to the fold line in the celestial sphere, the inferred POAM spectral asymmetry is about $10^{-5}$. 

As we have already seen from the wave behavior near a fold line or a cusp point, the generating functions are not proportional to $\omega$ along the control variables' direction, and the  amplitudes usually oscillate fast. This happens because a family of light-rays merge with each other at the optical caustic, and their interference gives these unusual frequency-dependences. It is  the same merging interference which also generates the net POAM shown here.

%-----------------------------------------------------------------------------------------------------------------------------------------------------------------------------------------------------------

\section{Conclusion}\label{sec5}

In this work, we have studied a wave that is emitted near a rotating black hole and eventually reaches a distant observer. We demonstrated that measuring the distortion of the wavefront provides an independent channel for obtaining information about the source, in addition to light-bending or other spectroscopic measurements. In order to characterize the wavefront distortion, we have adopted the previously-established POAM decomposition and we have applied it to the wave, which we have computed using the retarded Green function in Kerr space-time. While the POAM spectra of waves scattered by Schwarzschild black holes are shown to be always symmetric, any rotation in the host black hole generically generates asymmetry in the POAM spectrum. Since the symmetric part of the spectrum is degenerate with the tilt angle of the observation plane, we conclude that it is more likely to extract 
information about the source  from the asymmetric part.

 The resulting POAM spectrum strongly depends on the temporal coherence of the emission source and the sky-location of the receiver.
On the one hand, we find that the 
main contribution to the symmetric spectrum weight comes from the  phase-variation in the sky of the direct signal.
For generic receiver's locations, the magnitude of the square root of the weight, $\sqrt{w_{\pm 1}}$, is of order of $\eta$, which is roughly the ratio between the size of the telescope and the Airy disk size of the image, and it could be as large as $0.6$ for millimeter-wavelength sources near Sagittarius A*. If the receiver is located near a cusp point on the sky, we have shown,
using techniques developed in catastrophe optics, that $\sqrt{w_{\pm 1}} \sim \eta/(\omega M)^{1/4}$; if the receiver is near a fold line in the sky, we have 
shown that $\sqrt{w_{\pm 1}} \sim \eta$ instead.

On the other hand, we have shown that the asymmetric part of the POAM spectrum could be generated by the beating between phase and amplitude variation of the primary signal, or by the beating between the phase variation of the primary and secondary signals. This beating reflects the physical origin of nonzero POAM, either by coherently combining rays mis-aligned optical-axis, or by the interference of adjacent light bundles.
Nevertheless, we find that the spectral asymmetry is generally too weak to be measured. This is in part due to the large distances between Earth and astrophysical black holes, which serve as strong-gravity lenses or ``phase plates". The best candidate for detection
might be an extended region in the accretion disk
near Sagittarius A* such that  Earth lies near a fold line of its radiation. In this case, the spectral asymmetry for the $|l|=1$ mode can be as large as $10^{-5}$. However, even with advanced techniques available to cancel the effects from atmosphere turbulence, e.g. adaptive optics methods, this signal is still far below the sensitivity of current radio telescopes, such as millimeter-wavelength Very Long Baseline Interferometry arrays.

\acknowledgements
We thank Barry Wardell, Latham Boyle, Avery Broderick and Haixing Miao for many helpful discussions. HY thanks Yanbei Chen for introducing this project and providing valuable comments on the manuscript.
This research is funded by NSF Grants PHY-1068881 and PHY-1005655, CAREER Grants PHY-0956189 and PHY-1055103, NASA Grant No.NNX09AF97G, the Sherman Fairchild Foundation, the Brinson Foundation, and the David and Barbara Groce Startup Fund at Caltech. HY acknowledges supports from the Perimeter Institute for Theoretical Physics and the Institute for Quantum Computing. 
MC  thanks the Perimeter Institute for Theoretical Physics for hospitality and financial support.
Research at Perimeter Institute is supported by the Government of Canada and by the Province of Ontario though the Ministry of Research and Innovation.
 
\appendix


\begin{thebibliography}{99}
\bibitem{Tamm} C.\ Tamm and C.\ O.\ Weiss, J.\ Opt.\ Soc.\ Am.\ B {\bf 7}, 1034, (1990). 
\bibitem{Mair} A.\ Mair, A.\ Vaziri, G.\ Weihs, and A.\ Zeilinger, Nature {\bf 412}, 313, (2001).
\bibitem{Leach} J.\ Leach, M.\ J.\ Padgett, S.\ M.\ Barnett, S.\ Franke-Arnold and J.\ S.\  Courtial, \prl {\bf 88}, 257901, (2002).
\bibitem{Terriza} G.\ Molina-Terriza, J.\ P.\ Torres and L.\ Torner, \prl {\bf 88}, 013601, (2001).
\bibitem{Gibson} G.\ Gibson, J.\ Courtial, M.\ Padgett, M.\ Vasnetsov, V.\ Pasko,S.\ Barnett and S.\ F.\ Arnold, Opt. Exp. {\bf 12}, 5448, (2004).
\bibitem{Vinet} J.\ Y.\ Vinet, \prd {\bf 82}, 042003, (2010).
\bibitem{Harwitt} M.\ Harwitt, \apj {\bf 597}, 1266, (2003).
\bibitem{Tamburini} F.\ Tamburini, B.\ Thide, G.\ M.\ Terriza, G.\ Anzolin, Nature Physics {\bf 7}, 195, (2011).
\bibitem{Poisson} E.\ Poisson, A.\ Pound, and I.\ Vega, Living Review in Relativity {\bf 14}, 7, (2011).
\bibitem{Elias} N.\ M.\ Elias II, Astronomy and Astrophysics {\bf 492} 883, (2008).
\bibitem{Thorne} K.\ S.\ Thorne and R.\ D.\ Blandford {\it  Modern Classical Physics: Optics, Fluids, Plasmas, Elasticity, Relativity, and Statistical Physics}, Princeton University Press (2014).
\bibitem{Siegman} A.\ E.\ Siegman, {\it Lasers} (University Science Books, Sausalito, CA, 1986), Chap. 16.
\bibitem{Thide} B.\ Thide, H.\ Then, J.\ Sjoholm, K.\ Palmer, J.\ Bergman, T.\ D.\ Carozzi, Ya.\ N.\ Istomin, N.\ H.\ Ibragimov and R.\ Khamitova, \prl {\bf 99} 087701, (2007).
\bibitem{Rogers} A.\ E.\ E.\ Rogers, S.\ S.\ Doloeman, and J.\ M.\ Moran, \apj {\bf 109}, 1391, (1995).
\bibitem{Averycommu} Private communication with Avery Broderick.
\bibitem{Leaver:1986} E.\ W.\ Leaver, \prd{\bf 34}, 384 (1986).
\bibitem{Price:1972} R.\ H.\ Price, \prd {\bf 5}, 2419, (1972).
\bibitem{Ching} E.\ S.\ C.\ Ching, P.\ T.\ Leung, W.\ M.\ Suen and K.\ Young, \prd {\bf 52}, 2118 (1995).
\bibitem{Mino} Y.\ Mino, M.\ Sasaki, and T.\ Tanaka, \prd {\bf 55}, 3457 (1997).
\bibitem{WardellOtt} A.\ C.\ Ottewill and B.\ Wardell \prd {\bf 84}, 104039 (2011).
\bibitem{Casals2009} M.\ Casals, S.\ Dolan, A.\ C.\ Ottewill and B.\ Wardell, \prd {\bf 79}, 124044, (2009).
\bibitem{Dolan2011} S.\ R.\ Dolan and A.\ C.\ Ottewill \prd {\bf 84}, 104002 (2011).
\bibitem{Casals2012a} M.\ Casals and A.\ Ottewill, \prl {\bf 109}, 111101, (2012).
\bibitem{anil} A.\ Zenginoglu and C.\ R.\ Galley, \prd {\bf 86}, 064030, (2012).
\bibitem{Yang2013} H. Yang, F.\ Zhang, A.\ Zimmerman, and Y.\ Chen, \prd {\bf 89}, 064014, (2014).
\bibitem{Casals2009a} M.\ Casals, S.\ Dolan, A.\ C.\ Ottewill and B.\ Wardell, \prd {\bf 79}, 124043, (2009).
\bibitem{Casals2013} M.\ Casals, S.\ Dolan, A.\ C.\ Ottewill and B.\ Wardell, \prd {\bf 88}, 044022 (2013).
\bibitem{wardell2014self}  B.\ Wardell, C.\ R.\ Galley, A.\ Zenginoglu, M.\ Casals,  S.\ R.\ Dolan and A.\ C.\ Ottewill, arXiv:1401.1506
\bibitem{Casals:2012px} M.\ Casals and B.\ C.\ Nolan \prd {\bf 86}, 024038 (2012).
\bibitem{Hod} S.\ Hod,  \prl {\bf 84}, 10, (2000).
\bibitem{Casals2012b} M.\ Casals and A.\ Ottewill, \prd {\bf 87}, 064010, (2013).
\bibitem{Mashhoon} B.\ Mashhoon \prd {\bf 31}, 290 (1985).
\bibitem{Friedlander} F. G. Friedlander, {\it The Wave Equation on a Curved Space-time}, Cambridge University Press (1975).
\bibitem{Casals2012c} M.\ Casals and A.\ Ottewill, \prd {\bf 86}, 024021, (2012).
\bibitem{Yang2012a} H.\ Yang, D.\ A.\ Nichols, F.\ Zhang, A.\ Zimmerman, Z.\ Zhang and Y.\ Chen, \prd {\bf 86}, 104006, (2012).
\bibitem{Boyle} L.\ Boyle and M.\ Russo, arXiv:1110.2789.
\bibitem{mtw} C.\ W.\ Misner, K.\ S.\ Thorne, and J.\ A.\ Wheeler, {\it Gravitation} (San Francisco: W.\ H.\ Freeman and Co., 1973)
\bibitem{berry} M.\ V.\ Berry and C.\ Upstill, {\it Progress in Optics} {\bf 18}, (1980).
\bibitem{Bozza} V.\ Bozza, \prd {\bf 78}, 063014, (2008).
\bibitem{Rauch} K.\ Rauch and R.\ Blanford, \apj {\bf 421}, 46, (1994).
\end{thebibliography}
\end{document}